\documentclass[useAMS,usenatbib]{mnras}

\usepackage{verbatim} 
\usepackage{natbib} 
\usepackage{amsmath} 
\usepackage{amsbsy}
\usepackage{amssymb}
\usepackage{mathrsfs} 
\usepackage{lscape} 
\usepackage{graphicx}
\usepackage{epstopdf}
\usepackage{deluxetable}
\usepackage{ctable} 
\usepackage{fixltx2e} 
\usepackage{url}
\usepackage{txfonts}
\usepackage[T1]{fontenc} 
\usepackage{aecompl}

\newcommand{\acknowledgments}{\begin{small}\section*{Acknowledgments}\end{small}}

\newcommand{\hkpc}{h^{-1}{\rm kpc}}
\newcommand{\hmpc}{h^{-1}{\rm Mpc}}
\newcommand{\lcdm}{$\Lambda$CDM}
\newcommand{\Msun}{M$_{\odot}$}
\newcommand{\Msunh}{M$_{\odot}h^{-1}$}

\newcommand{\Mbulge}{$M_{\rm BH}$--$M_{\rm bulge}$}
\newcommand{\MM}{$M_{\rm BH}$--$M_{\star}$}

\newcommand{\pathI}{./fig}

\title[Gravitational torque-driven black hole growth and feedback]{Gravitational torque-driven black hole growth and feedback in cosmological simulations}

\author[D. Angl{\'e}s-Alc{\'a}zar, et al.]{
\parbox[t]{\textwidth}{\vspace{-1cm}
Daniel Angl{\'e}s-Alc{\'a}zar$^{1}$\thanks{E-mail: anglesd@northwestern.edu}, 
Romeel Dav{\'e}$^{2,3,4}$,
Claude-Andr{\'e} Faucher-Gigu{\`e}re$^{1}$,
Feryal {\"O}zel$^5$,
Philip F. Hopkins$^6$}\\
$^1$Center for Interdisciplinary Exploration and Research in Astrophysics (CIERA) and Department of Physics and Astronomy, \\ Northwestern University, 2145 Sheridan Road, Evanston, IL 60208, USA.\\
$^2$University of the Western Cape, Bellville, Cape Town 7535, South Africa.\\
$^3$South African Astronomical Observatories, Observatory, Cape Town 7925, South Africa.\\
$^4$African Institute for Mathematical Sciences, Muizenberg, Cape Town 7945, South Africa.\\
$^5$Astronomy Department, University of Arizona, Tucson, AZ 85721, USA.\\
$^6$TAPIR, Mailcode 350-17, California Institute of Technology, Pasadena, CA 91125, USA.}

\date{Submitted to MNRAS, March, 2016\vspace{-0.0cm}}
\pubyear{2016}

\begin{document}
\label{firstpage}
\maketitle

 \begin{abstract}

We investigate black hole--host galaxy scaling relations in cosmological simulations with a self-consistent black hole growth and feedback model. 
The sub-grid accretion model captures the key scalings governing angular momentum transport from galactic scales down to parsec scales, while our kinetic feedback implementation enables the injection of outflows with properties chosen to match observed nuclear outflows.
We show that ``quasar mode" feedback can have a large impact on the thermal properties of the intergalactic medium and the growth of galaxies and massive black holes for kinetic feedback efficiencies as low as 0.1\,\% relative to the bolometric luminosity.
Nonetheless, our simulations suggest that the black hole--host scaling relations are only weakly dependent on the effects of black hole feedback on galactic scales, owing to feedback suppressing the growth of galaxies and massive black holes by a similar amount.
In contrast, the rate at which gravitational torques feed the central black hole relative to the host galaxy star formation rate governs the slope and normalization of the black hole--host correlations.  
Our results suggest that a common gas supply regulated by gravitational torques is the primary driver of the observed co-evolution of black holes and galaxies.
 
 \end{abstract}

\begin{keywords}
galaxies: formation --- 
galaxies: evolution --- 
galaxies: active --- 
quasars: supermassive black holes ---
intergalactic medium ---
cosmology: theory\vspace{-0.5cm}
\end{keywords}

\section{Introduction}

\vspace{-0.1cm}
 
The energy released by accretion onto supermassive black holes may have a profound effect on the evolution of galaxies \citep{Silk1998,Somerville2008,Cattaneo2009}.  Indeed, contemporary models of galaxy formation appear to require feedback from active galactic nuclei (AGN) to suppress star formation in galaxies at high masses \citep{SomervilleDave2015}.
Recent years have seen increasing observational evidence for AGN feedback, from radio-emitting jets powered by slowly accreting black holes to powerful winds driven by quasars \citep{Fabian2012,Heckman2014}.  While the overall effect is still unclear, the energy and momentum inferred from observed fast nuclear outflows \citep[e.g.][]{Tombesi2013,Nardini2015} and galaxy-scale winds \citep[e.g.][]{Feruglio2010,Rupke2011,Sturm2011,Greene2012_QSOoutflow,Maiolino2012,Liu2013,Cicone2014,Harrison2014} suggest that AGN feedback may have a significant impact on the evolution of massive black holes as well, particularly during phases of rapid growth where most black hole mass is believed to assemble \citep{Soltan1982,YuTremaine2002}.  

The observed correlations between the mass of central supermassive black holes and various stellar properties of their host galaxies \citep[e.g.][]{Haring2004,Hopkins2007_BHplaneObs,Gultekin2009,Graham2013,McConnell2013,Woo2013} are often interpreted as indirect evidence for the effects of feedback from black hole accretion on galactic scales.
Analytic models show that the black hole--galaxy scaling relations can be explained under the assumption that black holes regulate their own growth by the efficient coupling of feedback at galactic scales \citep{Silk1998,King2003,Murray2005,King2015_review}.  In these models, black holes grow to a critical mass at which feedback is able to expel the remaining gas in the galaxy, inhibiting further accretion as well as star formation in the host galaxy.  This scenario has been extensively explored in numerical hydrodynamic simulations over the last decade, where ``sub-grid" models are introduced in order to incorporate black hole growth as well as the effects of feedback on galactic scales \citep[e.g.][]{DiMatteo2005,Hopkins2005_QSOevol,Springel2005_BHmodel,Hopkins2006,Hopkins2007_BHplane,Sijacki2007,Booth2009,Choi2012_BHmodel,Debuhr2012,Dubois2012,Rosas-Guevara2015,Sijacki2015,Steinborn2015}.  The success of these models in explaining many observables of galaxies and quasars has contributed to establishing a paradigm in which the observed connection between black holes and galaxies is driven by feedback from the black hole itself.

While feedback self-regulation represents an interesting possibility, the detailed physics and overall efficiency with which black hole driven outflows interact with the inflowing gas feeding the accretion disk remains poorly understood.  
Regardless of the effects of feedback, the rate at which gravitational torques transport angular momentum at galactic scales may be the limiting factor for fueling AGN \citep{Hernquist1989,Shlosman1989,Shlosman1990,Jogee2006,Escala2007,Hopkins2010_MultiScale,Cen2015}.  Owing to its simplicity, most black hole accretion prescriptions used in hydrodynamic simulations of galaxy formation are based on the spherical Bondi parameterization \citep{Bondi1944,Bondi1952}.  In the Bondi parametrization, the angular momentum of the inflowing gas is explicitly neglected.  
\citet{Hopkins2010_MultiScale} addressed the problem of AGN fueling by performing multiple nested galaxy-scale simulations of progressively higher resolution.  These simulations showed that non-axisymmetric perturbations to the stellar potential drive gas into shocks that dissipate energy and angular momentum, dominating the net torque on the gas component and driving gas inflows down to sub-parsec scales.  \citet{Hopkins2011_Analytic} derived an analytic accretion rate estimator that captures the key scalings found in the numerical simulations, while showing that the spherical Bondi parameterization systematically fails to reproduce the gas inflow rates.
In very gas-rich systems, efficient local fragmentation may provide additional mechanisms for angular momentum transport, including the scattering of dense gas clumps and gravitational instability-driven turbulence \citep{Levine2008,Bournaud2011,Hopkins2015_NuclearSims}.

Recently, \citet{Angles-Alcazar2013,Angles-Alcazar2015} proposed an alternative scenario to explain the black hole--galaxy scaling relations motivated by a careful examination of the implications of different black hole accretion models.  
By post-processing cosmological simulations with the analytic accretion model of \citet{Hopkins2011_Analytic},  \citet{Angles-Alcazar2013,Angles-Alcazar2015} showed that self-regulation by feedback processes may not be required when the physics of gravitational torques is appropriately captured at a sub-grid level.
Instead, the rate at which gravitational torques drive gas inflows down to sub-parsec scales relative to the host galaxy star formation rate modulates the long-term co-evolution of massive black holes and galaxies.  In this torque-limited growth scenario, black holes and galaxies evolve on average toward the observed scaling relations, regardless of the initial conditions, and with no need for mass averaging through mergers \citep{Peng2007,Hirschmann2010,Jahnke2011} or additional self-regulation processes.  While the non-linear effects of feedback are neglected in this scenario, outflows are invoked in order to provide a significant mass loss from galactic scales to accretion disk scales, or within the accretion disk, thereby strongly suppressing black hole growth.  Nonetheless, there is no need for direct coupling of inflows and outflows to regulate black holes in a non-linear feedback loop.

It is worth noting that the adoption of Bondi accretion in simulations automatically implies the need for self-regulation by feedback processes in order to reproduce the observed black hole--galaxy scaling relations.  Indeed, any model in which the accretion rate depends on black hole mass as $\dot{M}_{\rm BH} \propto M_{\rm BH}^{p}$, with $p>1$ as in Bondi ($p=2$), yields divergent evolution of black hole mass relative to changes in the initial conditions unless additional feedback mechanisms ensure strong self-regulation \citep{Angles-Alcazar2015}.
The amount of feedback injected in the simulation relative to the accretion rate is generally chosen such that strong self-regulation occurs and yields the scaling relations \citep{DiMatteo2005}.
It is thus critical to break the degeneracy between fueling and feedback and evaluate their relative roles in driving the observed black hole--galaxy scaling relations.

Here, we revisit the torque-limited growth scenario by performing cosmological hydrodynamic simulations that follow the evolution of massive black holes at the centers of galaxies.  We present a new implementation of AGN feedback coupled to black hole accretion driven by gravitational torques that self-consistently captures the effects of feedback at galactic scales.  
Cosmological simulations will rely heavily on sub-grid models of black hole growth and feedback for the foreseeable future, emphasizing the importance of developing and testing models motivated by higher resolution calculations.  In this work, we do not attempt to build a comprehensive galaxy formation model to make detailed comparisons to observations.  Our main goal is rather to investigate the implications of black hole feedback in simulations when adopting a physically motivated accretion model based on gravitational torques.  We thus deliberately limit the complexity of galaxy formation physics included in our simulations and perform numerical experiments with various combinations of black hole model parameters to investigate the relative roles played by accretion and galaxy-scale AGN feedback in driving the overall connection between massive black holes and galaxies.

We begin by summarizing the properties of our simulations in \S\ref{sec:sim}, including descriptions of the implementation of gravitational torque accretion and black hole driven outflows.  We present the \Mbulge~relation predicted by our fiducial simulation in \S\ref{sec:mm}, where we analyze its dependence on various accretion and feedback model parameters. The global effects of black hole feedback in our simulations are reported in \S\ref{sec:glob}.  We discuss the implications of our findings in \S\ref{sec:dis} and present our conclusions in \S\ref{sec:conc}. 
Different aspects of the numerical robustness of our simulations are discussed in Appendix~\ref{sec:appendix:rob}.

\begin{footnotesize}
\ctable[caption={{\normalsize Parameters of simulations}\label{tbl:sims}},center,star]{lcccccccccc}{
\tnote[ ]{
(1) Name: Simulation designation.\\
(2) $m_{\rm b}$: Initial baryonic particle mass (\Msunh).\\
(3) $\epsilon_{\rm b}$: Minimum baryonic force softening length ($\hkpc$).\\
(4) $M_{\rm seed}$: Physical mass of black hole seeds (\Msunh).\\
(5) $\epsilon_{\rm T}$: Normalization of $\dot{M}_{\rm Torque}$.\\
(6) $v_{\rm out}$: Assumed velocity of AGN-feedback driven outflows (km\,s$^{-1}$).\\
(7) $p_{\rm b}$: Total momentum flux of outflows in units of $L_{\rm bol}/c$.\\
(8) $z_{\rm end}$: Final simulation redshift.\\
(9) $\dot{M}_{\rm out}/\dot{M}_{\rm BH}$: Mass outflow rate relative to black hole accretion rate (determined by $v_{\rm out}$ and $p_{\rm b}$).\\
(10) $\epsilon_{\rm k}$: Kinetic feedback efficiency, defined as $\epsilon_{\rm k} \, \equiv \, \frac{1}{2} \dot{M}_{\rm out} v_{\rm out}^{2} \, / \, L_{\rm bol}$ (determined by $v_{\rm out}$ and $p_{\rm b}$).
}
}{
\hline\hline
\noalign{\vskip 0.5mm}
\multicolumn{1}{c}{Name} &
\multicolumn{1}{c}{$m_{\rm b}$} & 
\multicolumn{1}{c}{$\epsilon_{\rm b}$} &  
\multicolumn{1}{c}{$M_{\rm seed}$} &
\multicolumn{1}{c}{$\epsilon_{\rm T}$} &
\multicolumn{1}{c}{$v_{\rm out}$} & 
\multicolumn{1}{c}{$p_{\rm b}$} &
\multicolumn{1}{c}{$z_{\rm end}$} &
\multicolumn{1}{c}{notes} &
\multicolumn{1}{c}{$\dot{M}_{\rm out}/\dot{M}_{\rm BH}$} &
\multicolumn{1}{c}{$\epsilon_{\rm k}$}
 \\ 
\noalign{\vskip 0.2mm}
\hline
\noalign{\vskip 1mm}
{\bf n256-fid} &  6.4e6  &  0.16  &  $10^5$ &  0.5  &  $10^3$  & 1 & 0 & Fiducial simulation & 30 & 1.6e-3\\ 
{\bf n256s6}  &  6.4e6  &  0.16  &  $10^6$ &  0.5  &  $10^3$  & 1 & 0 & Over-massive seed & 30 & 1.6e-3\\ 
{\bf n256s4}  &  6.4e6  &  0.16  &  $10^4$ &  0.5  &  $10^3$  & 1 & 0 & Under-massive seed & 30 & 1.6e-3\\ 
{\bf n256eH} &  6.4e6  &  0.16  &  $10^5$ &  5  &  $10^3$  & 1 & 0 & High normalization & 30 & 1.6e-3\\  
{\bf n256eL}  &  6.4e6  &  0.16  &  $10^5$ &  0.05 &  $10^3$  & 1 & 0 & Low normalization & 30 & 1.6e-3\\  
{\bf n256v4}  &  6.4e6  &  0.16  &  $10^5$ &  0.5    &  $10^4$  & 1 & 0 & High velocity outflows & 3 & 1.6e-2\\  
{\bf n256p20} &  6.4e6  &  0.16  &  $10^5$ &  0.5  &  $10^3$  & 20 & 0 & Large momentum boost & 600 & 3.3e-2\\
{\bf n256-nf}  &  6.4e6  &  0.16  &  $10^5$ &  0.5  &  $0$  &  0  & 0 & No feedback & 0 & 0\\
{\bf n256-mfm} &  6.4e6  &  0.16  &  $10^5$ &  0.5    &  $10^3$  & 1 & 0 &  Meshless hydrodynamics & 30 & 1.6e-3\\  
{\bf n512}   &  8.0e5  &  0.08  &  $10^5$ &  0.5    &  $10^3$  &  1  &  2  & High resolution & 30 & 1.6e-3\\
\hline\hline\\
}
\end{footnotesize}

\section{Simulations}\label{sec:sim}

Our main simulations use the N-body + hydrodynamics simulation code GIZMO\footnote{\url{www.tapir.caltech.edu/~phopkins/Site/GIZMO.html}} \citep{Hopkins2015_Gizmo} in ``P-SPH" mode, a pressure-entropy formulation of smooth particle hydrodynamics (SPH) that minimizes the errors of previous SPH formulations regarding fluid mixing instabilities \citep{Hopkins2013_PSPH,Saitoh2013}.  Gravitational forces are computed using a modified version of the tree-particle-mesh algorithm of the GADGET-2 code \citep{Springel2005_Gadget}, including adaptive gravitational softenings following \citet{Price2007}.  We use a time-step limiter to handle strong black hole feedback events \citep{Durier2012}.

We include radiative cooling from primordial gas \citep{Katz1996}, metal-line cooling \citep{Wiersma2009}, and photoionization heating from an optically thin UV background \citep{Faucher-Giguere2009}.
Star formation is modeled following the sub-grid prescription of \citet{Springel2003_Multiphase}: gas particles with density $n_{\rm H} \gtrsim 0.13$\,cm$^{-3}$ are treated as a multi-phase fluid with cold clouds embedded in a hot medium \citep{McKee1977}, which gives them an ``effective pressure" larger than the thermal pressure, based on sub-grid supernovae heating.
Gas particles are converted into star particles with a probability based on a \citet{Schmidt1959} law, such that the resulting SFRs are in agreement with the observed \citet{Kennicutt1998} relation.
We do not include star formation-driven winds, implying that the global properties of our simulated galaxies do not necessarily match observations.  Nevertheless, our simplified setup allows us to study the implications of torque-limited black hole growth and feedback for a wide range of host galaxy properties.

We adopt a ``standard" flat \lcdm~cosmology with parameters $\Omega_{\rm \Lambda} = 0.69$, $\Omega_{\rm M} = 0.31$, $\Omega_{\rm b} = 0.05$, $h = 0.68$, $\sigma_{8} = 0.82$, and $n = 0.97$, consistent with \citet{PlanckCollab2015}.
Our main simulation runs evolve a $[20\,\hmpc]^3$ comoving volume down to $z = 0$ employing $256^3$ gas and $256^3$ dark matter particles with masses
$m_{\rm b} = 6.4 \times 10^6$\,\Msunh~and $m_{\rm DM} = 3.4
\times 10^7$\,\Msunh, respectively.  Cosmological initial conditions were generated using MUSIC \citep{Hahn2011}. The minimum comoving softening length is set to 2\,\% of the mean interparticle distance for dark matter particles, $\epsilon_{\rm DM} = 1.6\,h^{-1}$\,kpc, while it is allowed to decrease down to $\epsilon_{\rm b} = 0.16\,h^{-1}$\,kpc for baryonic particles (gas, stars, and black holes).  The minimum SPH smoothing lengths are comparable to or smaller than the minimum softening length $\epsilon_{\rm b}$.
We present a resolution convergence test in \S\ref{sec:appendix:res}. 
The black hole accretion and feedback models used here are appropriate for large box simulations:  our mass and force resolution are comparable to, e.g., the Illustris \citep{Vogelsberger2014} and EAGLE \citep{Schaye2015} simulations.

All runs use the same basic simulation parameters with the exception of quantities specific to black hole accretion and feedback, which are varied as described below.   
Moreover, we evaluate the robustness of our results with respect to the hydrodynamics solver in \S\ref{sec:appendix:mfm}, where we employ the Lagrangian Godunov-type ``meshless finite mass" (MFM) method \citep{Hopkins2015_Gizmo}.
The simulation suite presented in this paper is summarized in Table~\ref{tbl:sims}. 
Additional runs not included here were performed to test various numerical aspects of our simulations (fixed versus adaptive softening lengths, isotropic versus collimated outflows) as well as to expand the range of black hole parameters ($M_{\rm seed}$, $\epsilon_{\rm T}$, $v_{\rm out}$, $p_{\rm b}$; see below).  The conclusions presented in this paper are not affected by any of the effects investigated in these tests, so we do not show them for brevity.

In the remainder of this section, we describe how we model black hole seeding (\S\ref{sec:seeds}), black hole dynamics and mergers (\S\ref{sec:merg}), black hole accretion (\S\ref{sec:acc}), and black hole feedback (\S\ref{sec:feedback}) in our simulations, as well as how we perform our main analysis (\S\ref{sec:analysis}).

\subsection{Black hole seeds}\label{sec:seeds}

Several black hole seed production mechanisms have been proposed, including the formation of black holes with mass $\sim 10^{2}$\,\Msun~as remnants of population III stars \citep[e.g.][]{Madau2001} and the formation of black holes with masses as large as $\sim 10^{5-6}$\,\Msun~by the direct collapse of gas in pre-galactic halos \citep[e.g.][]{Begelman2006,Choi2013_MassiveSeeds}.  Despite much recent work, major uncertainties remain on the nature of black hole seeds \citep[for a review, see][]{Volonteri2010}.

For simplicity, we do not attempt to mimic the physics of any seed formation mechanism in detail and simply assume that there is one black hole located at the center of each galaxy when it is first resolved in the simulation.  We specify the initial mass of the black hole ($M_{\rm seed}$) as well as the minimum galaxy stellar mass allowed to host a seed, $M_{\star}^{\rm min} = \gamma_{\rm BH} \times M_{\rm seed}$.
Our approach follows \citet{DiMatteo2008}.  We use a Friends-of-Friends (FOF) algorithm to identify dark matter halos during the simulation.  If the FOF group does not already contain a black hole particle and its stellar mass is $M_{\star}^{\rm fof} > \gamma_{\rm BH} \times M_{\rm seed}$, the gas particle with the highest density is converted into a black hole particle.

For our fiducial simulations, we employ $M_{\rm seed} = 10^5$\,\Msunh~and $\gamma_{\rm BH} = 10^3$ ($\gtrsim 15$ star particles), which places black holes and galaxies approximately on the local \Mbulge~relation, but we vary this.

\subsection{Black hole dynamics and mergers}\label{sec:merg}

At our typical resolution, the mass of black hole particles may be comparable to or smaller than the mass of dark matter particles.  The gravitational dynamics of individual black holes is thus not accurately captured in the simulation.
Following previous work \citep[e.g.][]{Springel2005_BHmodel,Booth2009,Sijacki2015}, we effectively fix the position of black hole particles to the location of the most bound particle (gas or star) within the radial aperture $R_{0}$ used to compute the accretion rate estimate.  Black hole particles are repositioned at every time step, provided that the relative velocity of the nearby most bound particle is lower than their mutual escape velocity.  Different improvements to the treatment of black hole dynamics have been proposed \citep[e.g.][]{Wurster2013} but this scheme is sufficient to compute the mass growth of ``well-behaved" central black holes in our simulations. 
We thus assume that dynamical friction is efficient enough to maintain black holes close to the center of galaxies but note that this may not properly capture the orbital decay of black holes in low mass galaxies and/or at high redshift \citep{Tremmel2015_DynFriction}.

Galaxy merger remnants will inevitably contain two or more massive black holes that may eventually merge, but our simulations lack the resolution to follow this process in detail.
Following \citet{Springel2005_BHmodel}, we simply allow any two black holes to merge instantaneously when they are located within $R_{0}$ if their relative velocity is lower than their mutual escape velocity.
We neglect the effects of gravitational recoils \citep{Blecha2011,Sijacki2011,Blecha2016}.

\subsection{Gravitational torque-driven accretion}\label{sec:acc}

Accretion rates are computed based on the gravitational torque model of \citet{Hopkins2011_Analytic}, which provides an estimate of the gas inflow rate $\dot{M}_{\rm Torque}$ driven by gravitational instabilities from galactic scales down to the accretion disk surrounding the central black hole:
\begin{equation}\label{eq:mdot} 
\dot{M}_{\rm BH} = (1 - \eta) \times \, \dot{M}_{\rm Torque},
\end{equation}
where we adopt a constant radiative efficiency $\eta = 0.1$  \citep[e.g.][]{YuTremaine2002,Marconi2004}.  
We allow black holes to exceed the Eddington accretion rate by up to a factor of 10.  This limit is rarely reached in our simulations.  We estimate $\dot{M}_{\rm Torque}$ based on properties of the host galaxy within a distance $R_{0}$ of each black hole \citep{Hopkins2011_Analytic}:

\begin{align}\label{eq:torque} 
\dot{M}_{\rm Torque} \; \approx \; \epsilon_{\rm T} \, f_{\rm d}^{5/2} \times 
\left ( \frac{M_{\rm BH}}{10^{8}\,{\rm M_{\odot}}} \right )^{1/6} 
\left ( \frac{M_{\rm d}(R_{0})}{10^{9}\,{\rm M_{\odot}}} \right ) \nonumber &\\
\times \left ( \frac{R_{0}}{100\,{\rm pc}} \right )^{-3/2}  \left (1 +
\frac{f_{0}}{f_{\rm gas}} \right )^{-1} \, {\rm M_{\odot}\,yr^{-1}}, &
\end{align}  
where $f_{\rm d}$ represents the disk fraction for the combined gas and stellar disk mass $M_{\rm d}(R_{0})$ within $R_{0}$, 
\begin{equation} 
f_{\rm d} \equiv M_{\rm d}(R_{0}) / (M_{\rm gas}(R_{0})+M_{\rm star}(R_{0})), 
\end{equation}
while $M_{\rm gas}(R_{0})$ and $M_{\rm star}(R_{0})$ represent the total gas and stellar masses within $R_{0}$.  Here, $f_{\rm gas}$ represents the gas fraction relative to the disk mass, 
\begin{equation}\label{eq:fgas}
f_{\rm gas} \equiv M_{\rm gas}(R_{0})/M_{\rm d}(R_{0}),
\end{equation} 
and
\begin{equation} f_{0} \approx 0.31 \, f_{\rm d}^{2} \, (M_{\rm d}(R_{0})/10^{9}{\rm M_{\odot}})^{-1/3}.
\end{equation}

The normalization factor $\epsilon_{\rm T}$ is intended to capture processes that affect the radial transport of gas on unresolved scales\footnote{Note that $\epsilon_{\rm T} \equiv \epsilon_{\rm m} \times \alpha_{\rm T}$ in the notation adopted by \citet{Angles-Alcazar2013,Angles-Alcazar2015}, where $\epsilon_{\rm m}$ is the ``mass retention rate" and $ \alpha_{\rm T}$ is the original normalization of $\dot{M}_{\rm Torque}$ in \citet{Hopkins2011_Analytic}.}, 
such as star formation, feedback from stars and the central black hole, and mass loss in winds from the accretion disk, which were not modeled explicitly in \citet{Hopkins2011_Analytic}.
The aperture $R_{0}$ is the distance enclosing 256 gas particles, with an upper limit of $2\,\hkpc$ (comoving) imposed throughout the simulation.
Evaluating equation (\ref{eq:torque}) requires separating the spheroidal and disk components of the galaxy center, which we do by means of the same kinematic decomposition as \citet{Angles-Alcazar2013,Angles-Alcazar2015}.
This method is described in \S\ref{sec:appendix:mbulge} in the context of galaxy bulge-disk decompositions.
Numerically, black hole accretion proceeds stochastically as in \citet{Springel2005_BHmodel}.  Gas particles within $R_0$ can get a fraction $f_{\rm m}$ of their mass subtracted (added to the black hole) with a probability that statistically satisfies the continuous mass growth given by equation~(\ref{eq:mdot}). 
A time step limiter is imposed on black hole particles such that black holes do not grow by more than 0.1\,\% of their current mass in a single time step.

\subsection{Black hole feedback}\label{sec:feedback}

We model AGN-driven outflows by stochastically kicking particles around the black hole with velocity $v_{\rm out}$,  with probability
\begin{equation}\label{eq:stoch2} 
p_{j} \; = \; \frac{ 1 - f_{\rm m}}{f_{\rm m}} \times \frac{w_{j}}{m_j} \times \dot{M}_{\rm BH} \, \Delta t,  
\end{equation}
where $w_{j}$ is a kernel weight ($\Sigma_j \, w_j = 1$) and $f_{\rm m}$ is the fraction of gas mass accreted by the black hole and subtracted from the gas particle before ejection.
This gives an outflow mass-loading $\dot{M}_{\rm out} / \dot{M}_{\rm BH} = (1 - f_{\rm m}) / f_{\rm m}$.
The ``momentum loading" and ``energy loading" trivially follow:
\begin{equation}\label{eq:facc}
p_{\rm b}  \, \equiv \, \frac{\dot{P}_{\rm out}}{L_{\rm bol}/c}  \, = \, \frac{v_{\rm out}}{\eta \,c} \, \left ( \frac{1 - f_{\rm m}}{f_{\rm m}} \right ),
\end{equation}
\begin{equation}\label{eq:ekin}
 \epsilon_{\rm k} \, \equiv \, \frac{\dot{E}_{\rm out}}{L_{\rm bol}} \, = \, \frac{1}{2\,\eta} \left ( \frac{v_{\rm out}}{c} \right )^{2} \left ( \frac{1 - f_{\rm m}}{f_{\rm m}} \right ),
\end{equation} 
where $L_{\rm bol} = \eta \, \dot{M}_{\rm BH} \, c^2$ and $c$ is the speed of light.

This is similar to kinetic wind implementations used in galactic nucleus \citep{Hopkins2015_NuclearSims}, galaxy merger \citep{Choi2012_BHmodel,Debuhr2012}, and cosmological ``zoom-in" simulations \citep{Choi2015_CosmoSim}.  The appropriate momentum and energy loading of winds depend on the physical scale \citep[e.g.][]{Faucher-Giguere2012_WindModel}.
Observed properties at different radii are uncertain, so we vary $p_{\rm b}$ and $v_{\rm out}$ (see Table 1).
Velocity kicks are directed radially from the black hole.  We also tested a model for collimated outflows, with kicks always in the direction of the angular momentum within $R_{0}$, but we found no significant differences in quantities studied here.  Similar star formation suppression efficiencies and black hole mass to galaxy mass ratios were obtained regardless of the outflow geometry.

\begin{figure}
\begin{center}
\includegraphics[scale=0.55]{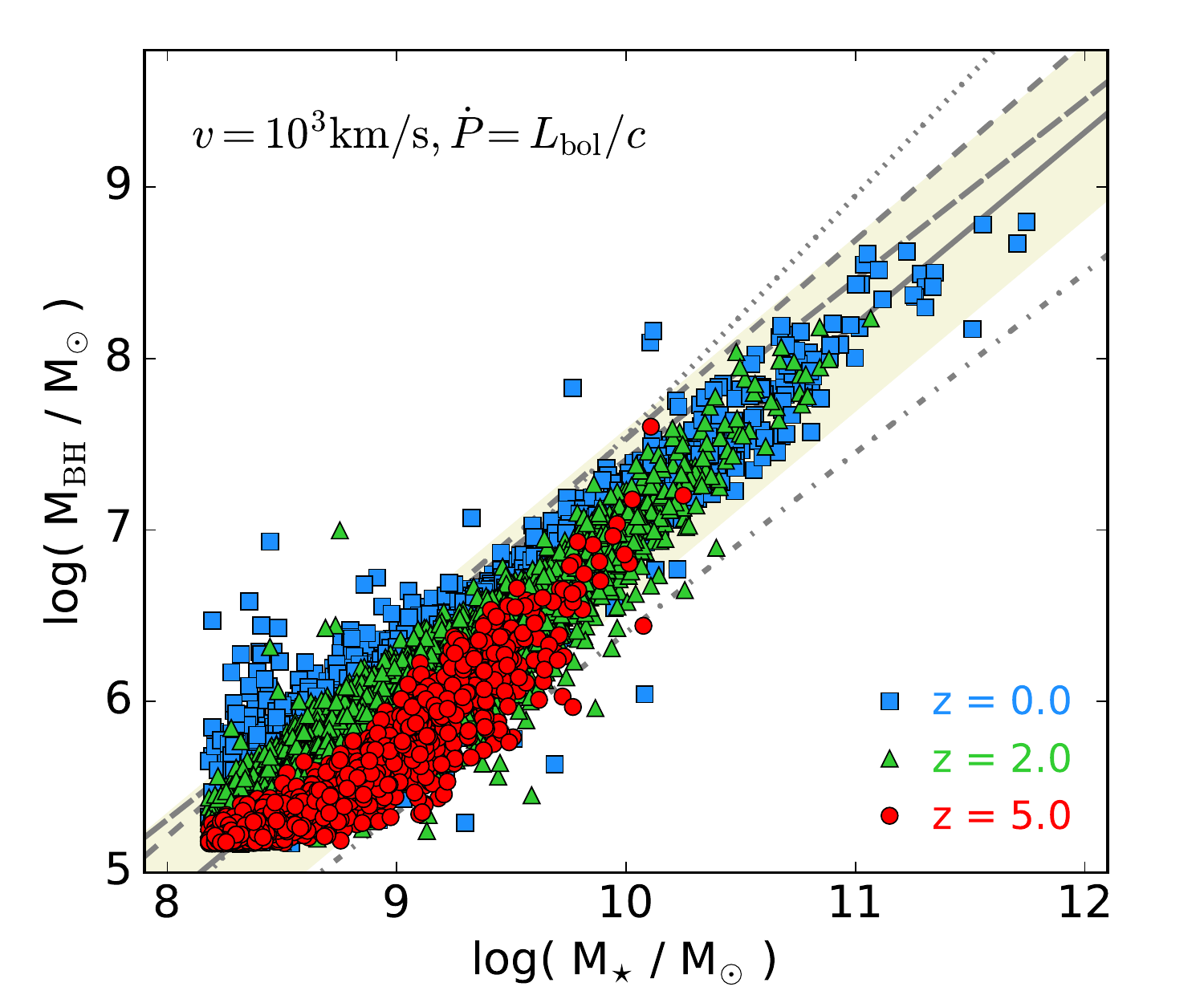}
\end{center}
\caption{\MM~relation at redshifts $z=5$ (red circles), $z=2$ (green triangles), and $z=0$ (blue squares) for our fiducial simulation (n256-fid).  
Gray lines indicate the observed \Mbulge~relations of \citet{Haring2004} (solid), \citet{McConnell2013} (long dashed), and \citet{Kormendy2013} (dashed), and the \MM~relations of \citet{Reines2015} for AGN (dot-dashed) and inactive (dotted) samples.  The beige shaded area corresponds to 0.5 dex scatter in $M_{\rm BH}$ relative to \citet{Haring2004}.
Our fiducial model (seed mass $M_{\rm seed} = 10^5$\,\Msunh, with quasar mode feedback and torque-limited accretion) agrees well with the observed $z=0$ relation and predicts very weak redshift evolution.}
\label{fig:fig1}
\end{figure}

\begin{figure}
\begin{center}
\includegraphics[scale=0.55]{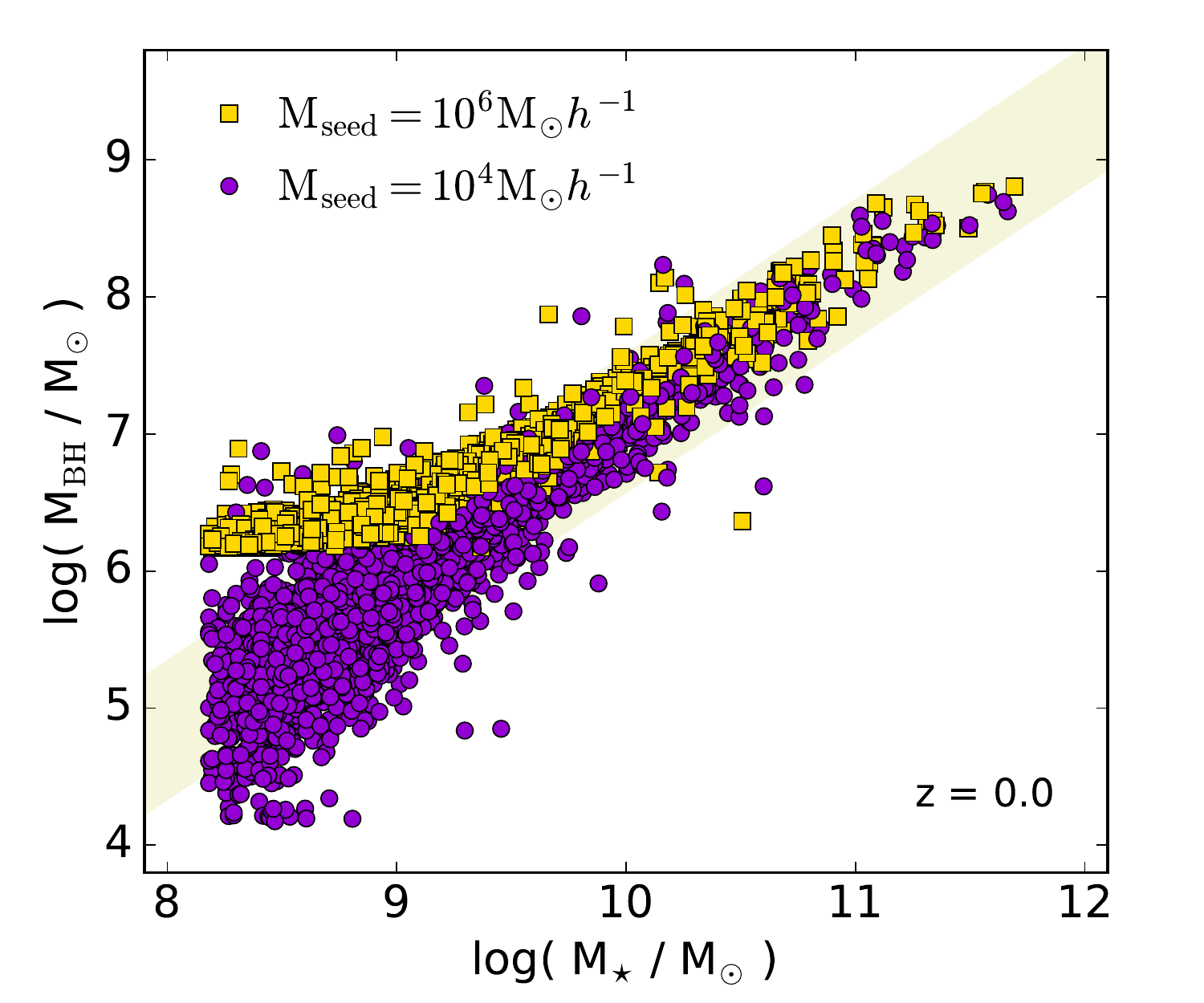}
\end{center}
\caption{Effects of the black hole seed mass on the \MM~relation at $z=0$.
Black holes with initial mass $M_{\rm seed} = 10^6$\,\Msunh (yellow squares) and $M_{\rm seed} = 10^4$\,\Msunh (purple circles) are seeded in galaxies $M_*  > 10^8$\,\Msunh.
The beige shaded area corresponds to 0.5 dex scatter in $M_{\rm BH}$ relative to \citet{Haring2004}.
Because the gravitational torque accretion rate is nearly independent of $M_{\rm BH}$, black hole seeds quickly converge to the same relation.}
\label{fig:fig2}
\end{figure}

\subsection{Analysis of simulations}\label{sec:analysis}

We identify dark matter halos at each redshift snapshot by means of the Amiga Halo Finder \citep{Gill2004_AHF,Knollmann2009_AHF}, using the evolving virial overdensity definition of \citet{Bryan1998}.
Galaxies are identified independently of their parent halos as gravitationally bound collections of gas and star particles by means of {\sc skid}\footnote{\url{http://www-hpcc.astro.washington.edu/tools/skid.html}.  We use a modified version provided as part of the SPHGR package \citep{Thompson2015_SPHGR}.}, where we impose a minimum density threshold for gas particles $\rho \geq 0.1$\,cm$^{-3}$ \citep[e.g.][]{Keres2005}.  
Galaxies may contain more than one black hole shortly after a galaxy merger and before the central black holes merge as well.  When comparing our simulations against the observed local \Mbulge~relation, we add up the masses of all black holes located within the stellar effective radius ($R_{\rm e}$), which we plot against the stellar mass contained within $R_{\rm e}$ for each galaxy ($M_{\star}$).  For simplicity, we do not attempt to compute the bulge mass of galaxies in analogy with observations, but replace it instead by $M_{\star}$ throughout this paper.  This facilitates comparisons to other numerical studies using $M_{\star}$ \citep[e.g.][]{Sijacki2015} as well as to observational studies at higher redshifts, where bulge masses are difficult to estimate.  We evaluate the implications of different definitions of host galaxy bulge mass in \S\ref{sec:appendix:mbulge}, where we show that our results depend only weakly on the exact definition and our main conclusions remain unchanged.

\section{The \MM~relation}\label{sec:mm}

Figure~\ref{fig:fig1} shows the \MM~relation obtained at different redshifts for our fiducial simulation, where we indicate the location of individual black holes and galaxies at $z=5$ (red circles), $z=2$ (green triangles), and $z=0$ (blue squares).  Our fiducial simulation is in good agreement with various observational determinations of the \Mbulge~relation in the local universe, indicated by the gray lines \citep{Haring2004,Kormendy2013,McConnell2013}. 
For comparison, we also show the observed \MM~relations of \citet{Reines2015} for AGN and inactive galaxies separately, where $M_{\rm BH}$ is related to the total stellar mass of galaxies as opposed to the bulge mass. 
Black holes with initial mass $M_{\rm seed} = 10^5$\,\Msunh~were placed at the center of galaxies as they first reached ${\rm M}_{\star} \approx 10^3\,{\rm M}_{\rm seed}$ in the simulation (i.e., roughly in agreement with the local \Mbulge~relation) and evolved on average along the scaling relation from early times down to $z=0$.
As we show in \S\ref{sec:eps}, the accretion rate normalization $\epsilon_{\rm T}$ governs the normalization of the scaling relation, while the slope arises naturally in the simulation as a consequence of the proportional growth of black holes and galaxies over cosmological timescales.  
Our fiducial simulation adopts $\epsilon_{\rm T} = 0.5$, a factor $\sim 10$ lower than estimated from simulations without black hole feedback in \citet{Hopkins2011_Analytic}.  This is similar within a factor of 2 to the normalization required in \citet{Angles-Alcazar2013,Angles-Alcazar2015} to match the local \Mbulge~relation by post-processing simulations performed with a cosmological code including different stellar feedback models \citep{Oppenheimer2008,Dave2013}.

\begin{figure}
\begin{center}
\includegraphics[scale=0.55]{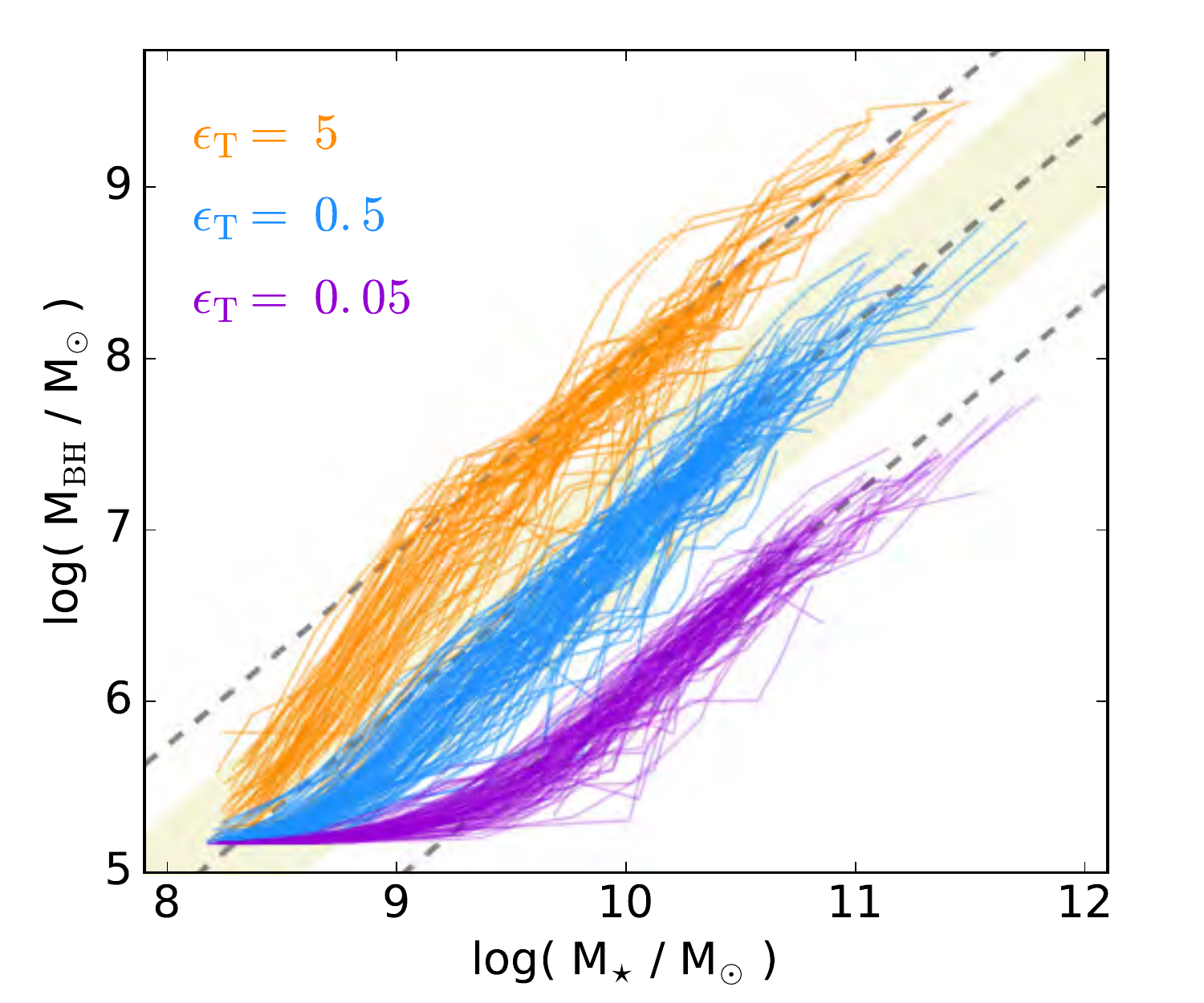}
\end{center}
\caption{Effects of the black hole accretion rate normalization on the \MM~relation.  We show evolutionary tracks for the 100 most massive black holes (on the \MM~plane at $z=0$) for $\dot{M}_{\rm BH} \propto \epsilon_{\rm T} = 5$, 0.5, 0.05.  Gray dashed lines indicate the \citet{Haring2004} relation and changing its normalization up and down by a factor of 10.  The accretion rate normalization $\epsilon_{\rm T}$ sets the normalization of the predicted \MM~relation,
with $M_{\rm BH}(M_*) \propto \epsilon_{\rm T}$.}
\label{fig:fig3}
\end{figure}

\begin{figure}
\begin{center}
\includegraphics[scale=0.27]{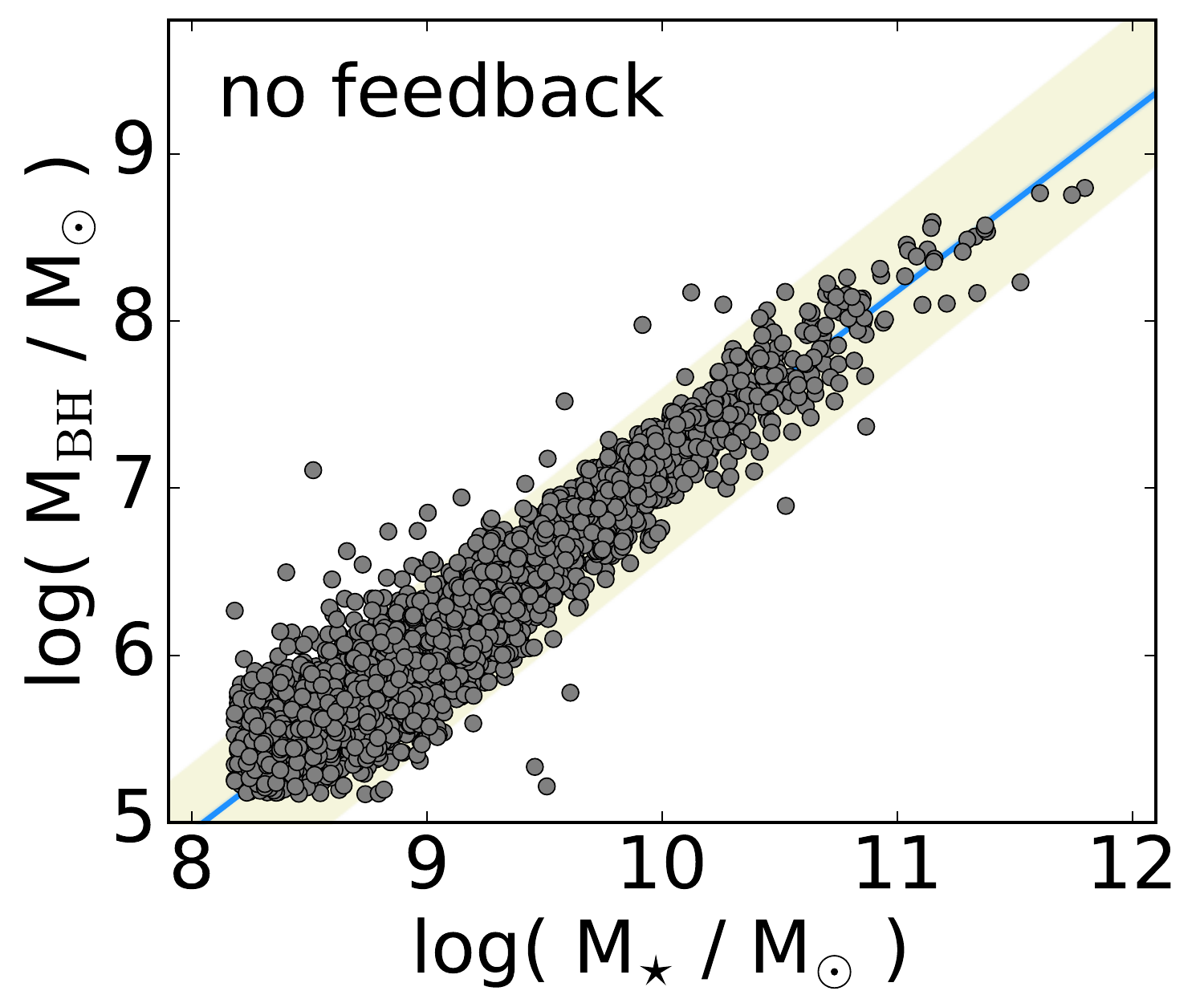}
\includegraphics[scale=0.27]{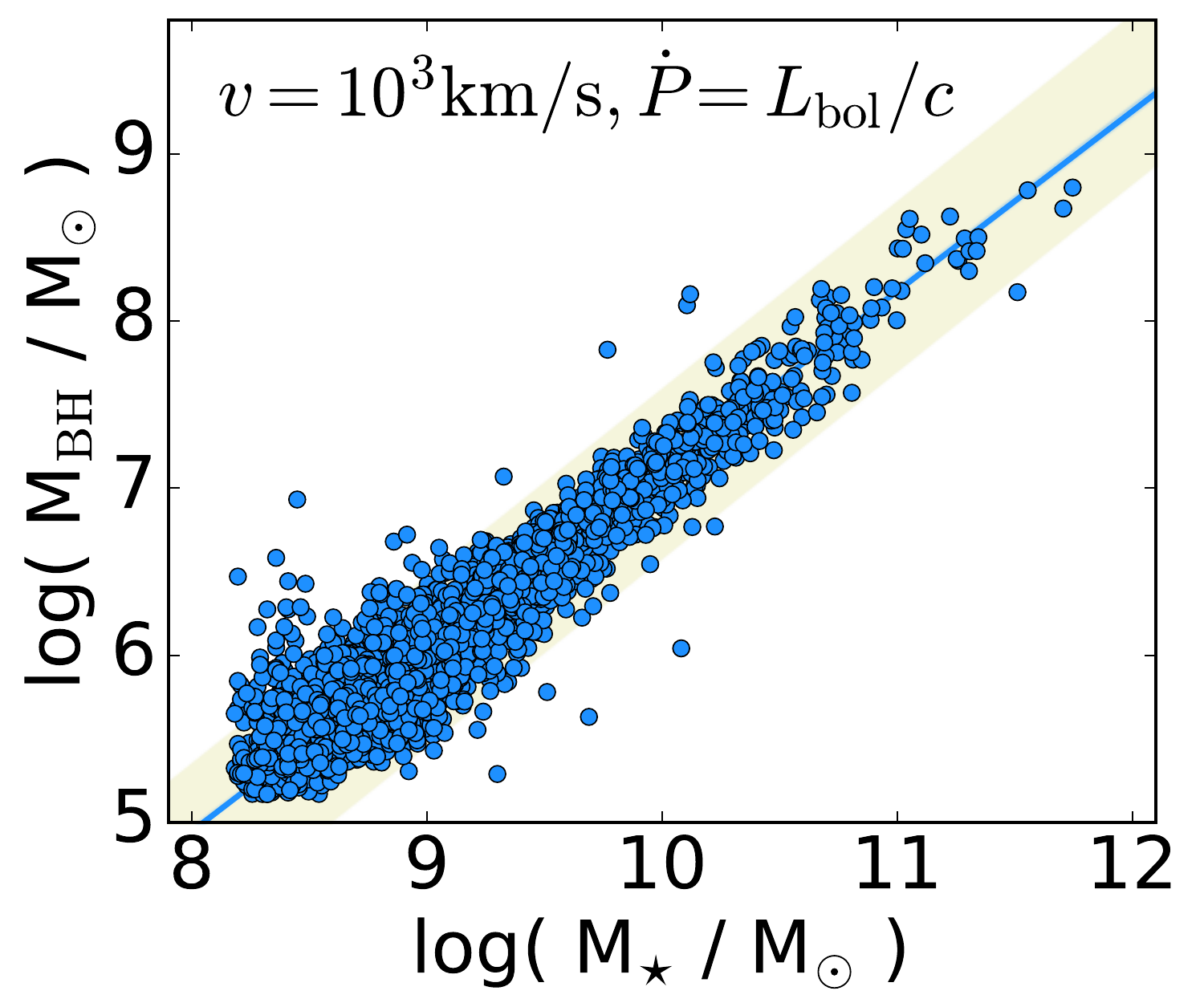}
\includegraphics[scale=0.27]{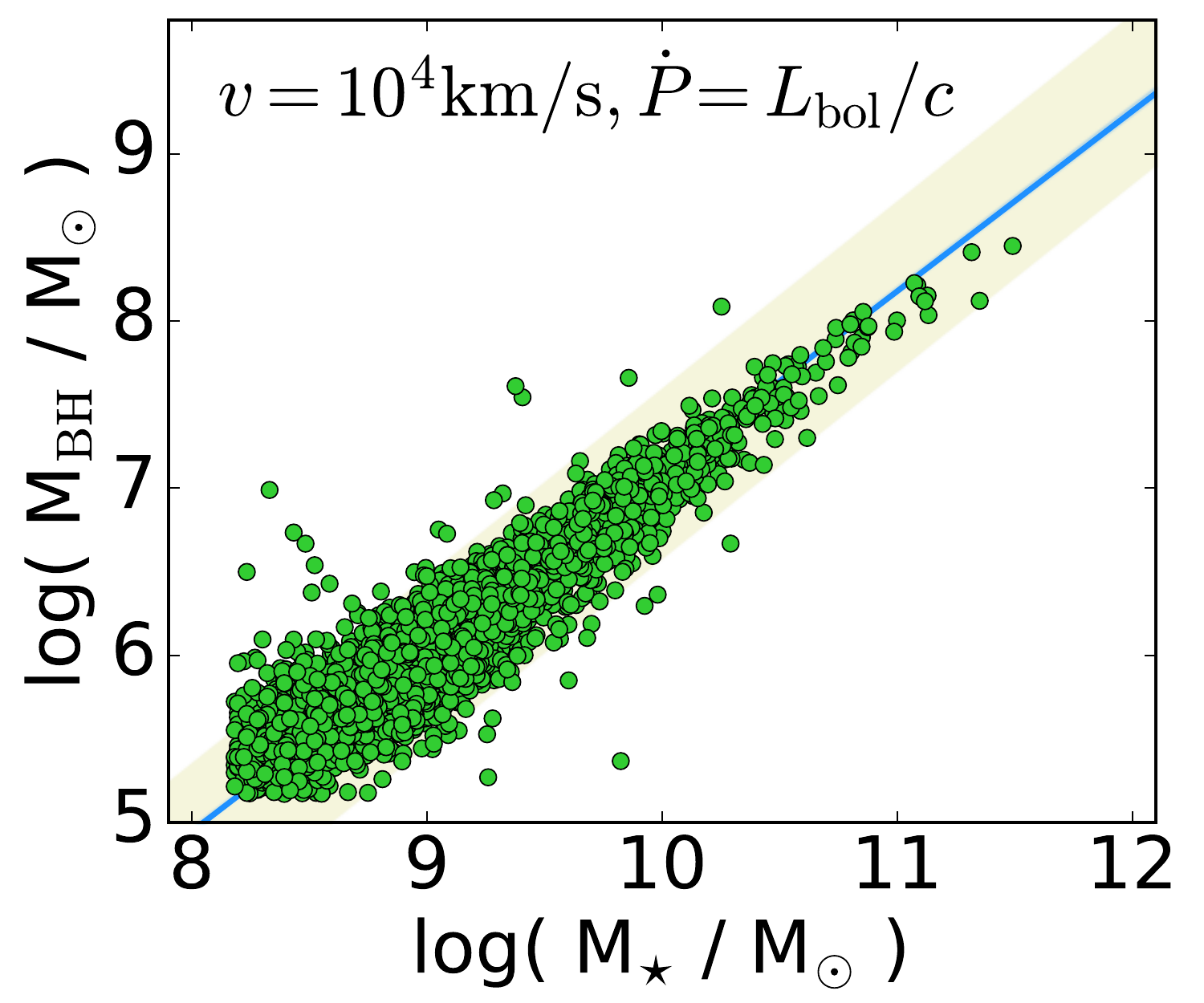}
\includegraphics[scale=0.27]{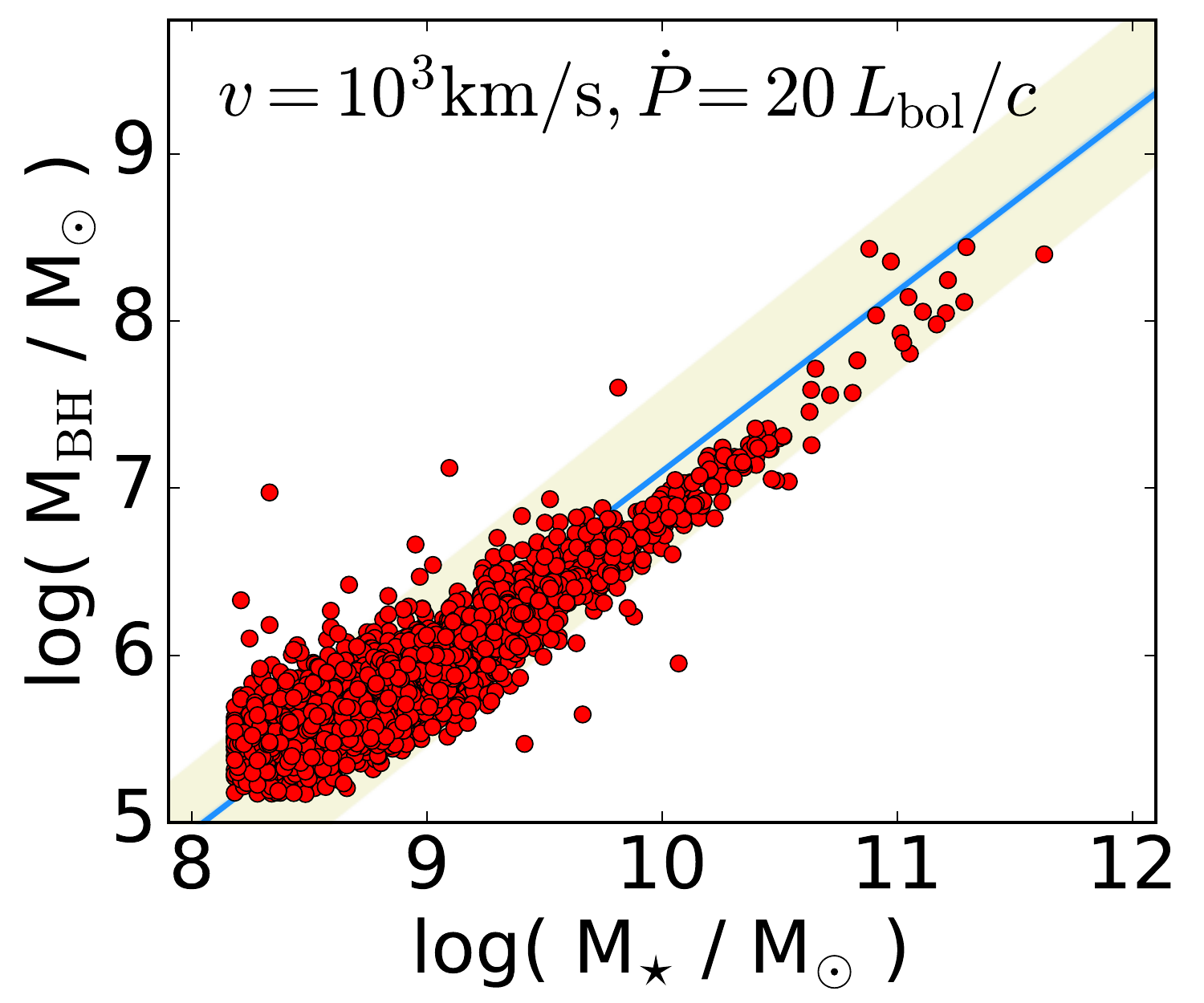}
\end{center}
\caption{Effects of galaxy-scale black hole feedback on the \MM~relation.
We show the scaling relation obtained at $z=0$ for simulations using different velocity ($v_{\rm out}$) and/or total momentum flux ($\dot{P}_{\rm out}$) for AGN-driven outflows coupled on $\sim$kpc scales, including (i) no explicit treatment of black hole feedback (top left), (ii) $v_{\rm out} = 10^3$\,km\,s$^{-1}$ and $\dot{P}_{\rm out} = L_{\rm bol}/c$ (top right; fiducial simulation), (iii) $v_{\rm out} = 10^4$\,km\,s$^{-1}$ and $\dot{P}_{\rm out} = L_{\rm bol}/c$ (bottom left), and (iv)  $v_{\rm out} = 10^3$\,km\,s$^{-1}$ and $\dot{P}_{\rm out} = 20\,L_{\rm bol}/c$ (bottom right).  All simulations use the same black hole accretion parameters.
Blue solid lines indicate the best power law fit to our fiducial \MM~relation.  
The beige shaded area corresponds to 0.5 dex scatter in $M_{\rm BH}$ relative to \citet{Haring2004}.
Although different galaxy-scale feedback choices affect both $M_*$ and $M_{\rm BH}$ significantly, they move along the same \MM~relation.}
\label{fig:fig4}
\end{figure}

We compute the best power-law fit to the \MM~relation at each redshift as
\begin{equation}\label{eq:fit}
{\rm log}_{10} \left ( \frac{M_{\rm BH}}{M_{\odot}} \right ) = \alpha + \beta \; {\rm log}_{10} \left ( \frac{M_{\star}}{10^{10}M_{\odot}} \right ),
\end{equation}
where only galaxies with $M_{\star} > 10^{9.5}$\,\Msun~are included, to exclude low mass black holes affected by our seed mass choice.
Our fiducial simulation yields $\alpha \approx 7.1$ and $\beta \approx 1.1$ at $z=0$ and we see no significant evolution in the normalization or slope of the relation.
Various observations differ regarding the normalization of the \Mbulge~relation.  Since we show this is determined by $\epsilon_{\rm T}$ in these models (\S\ref{sec:eps}), we can adjust our prediction systematically by changing $\epsilon_{\rm T}$.  In the following, we adopt the \citet{Haring2004} relation as our fiducial reference.

\subsection{Effects of different seed masses}\label{sec:ini}

Figure~\ref{fig:fig2} shows the \MM~relation at $z=0$ for two simulations, where we introduce black hole seeds with mass $M_{\rm seed} = 10^4$\,\Msunh~or $M_{\rm seed} = 10^{6}$\,\Msunh~in galaxies with stellar mass $M_* > 10^8$\,\Msunh~(i.e. $10\times$ smaller/larger than our default).
Despite initial masses differing by two orders of magnitude, Figure~\ref{fig:fig2} shows that black holes converge onto the local relation by $z=0$.  This is because $\dot{M}_{\rm Torque}$ is nearly independent of black hole mass (equation~\ref{eq:torque}), so that the proportional growth rate $\dot{M}_{\rm BH} / M_{\rm BH} \; \propto \; M_{\rm BH}^{-5/6}$ of under-massive (over-massive) black holes is faster (slower) relative to black holes lying on the \MM~relation for the same host galaxies.  The timescale for convergence onto the local relation depends on the initial $M_{\rm seed}/M_*$ ratio and redshift \citep{Angles-Alcazar2015}, after which black holes lose memory of their seed mass.

\begin{figure*}
\begin{center}
\includegraphics[scale=0.62]{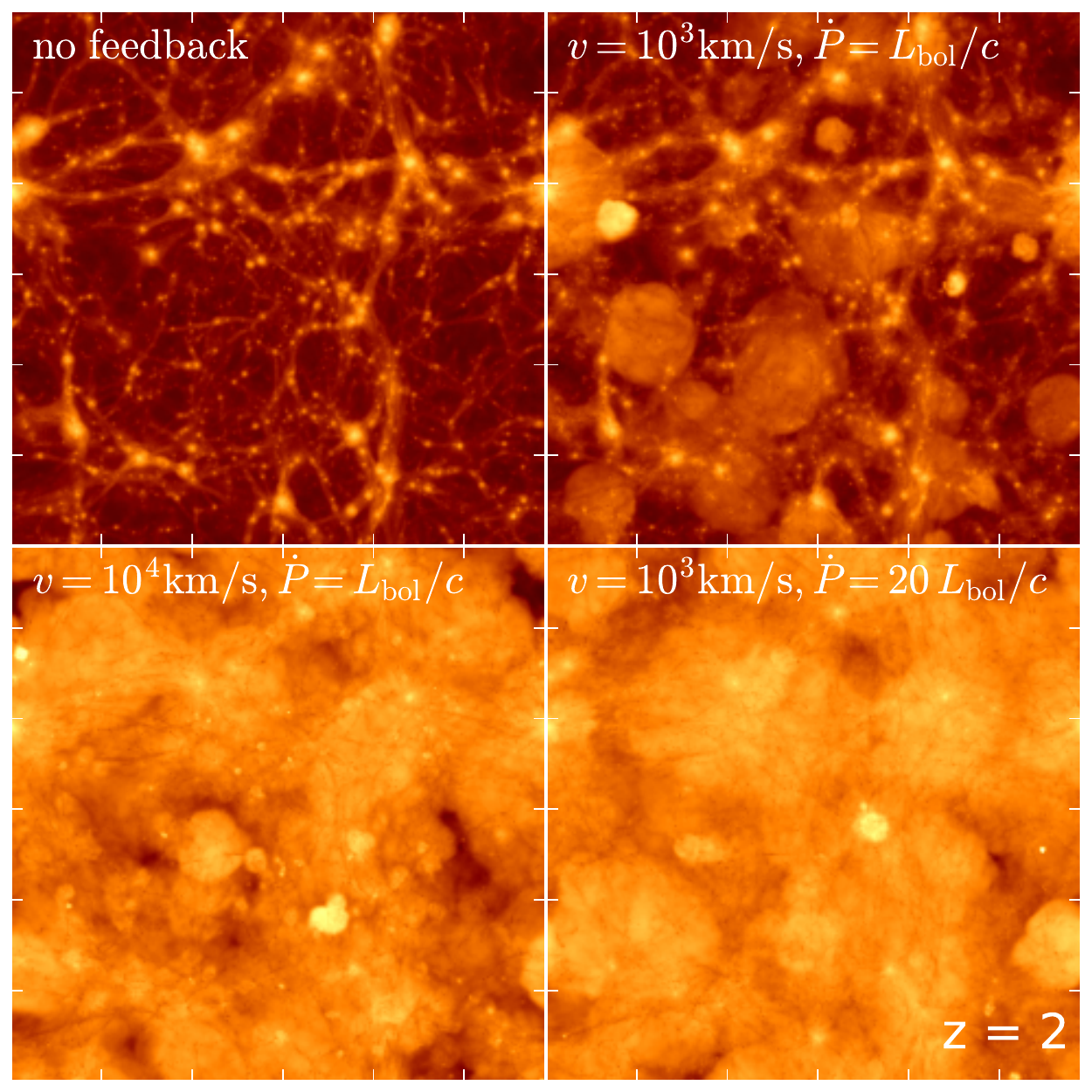}
\includegraphics[scale=0.62]{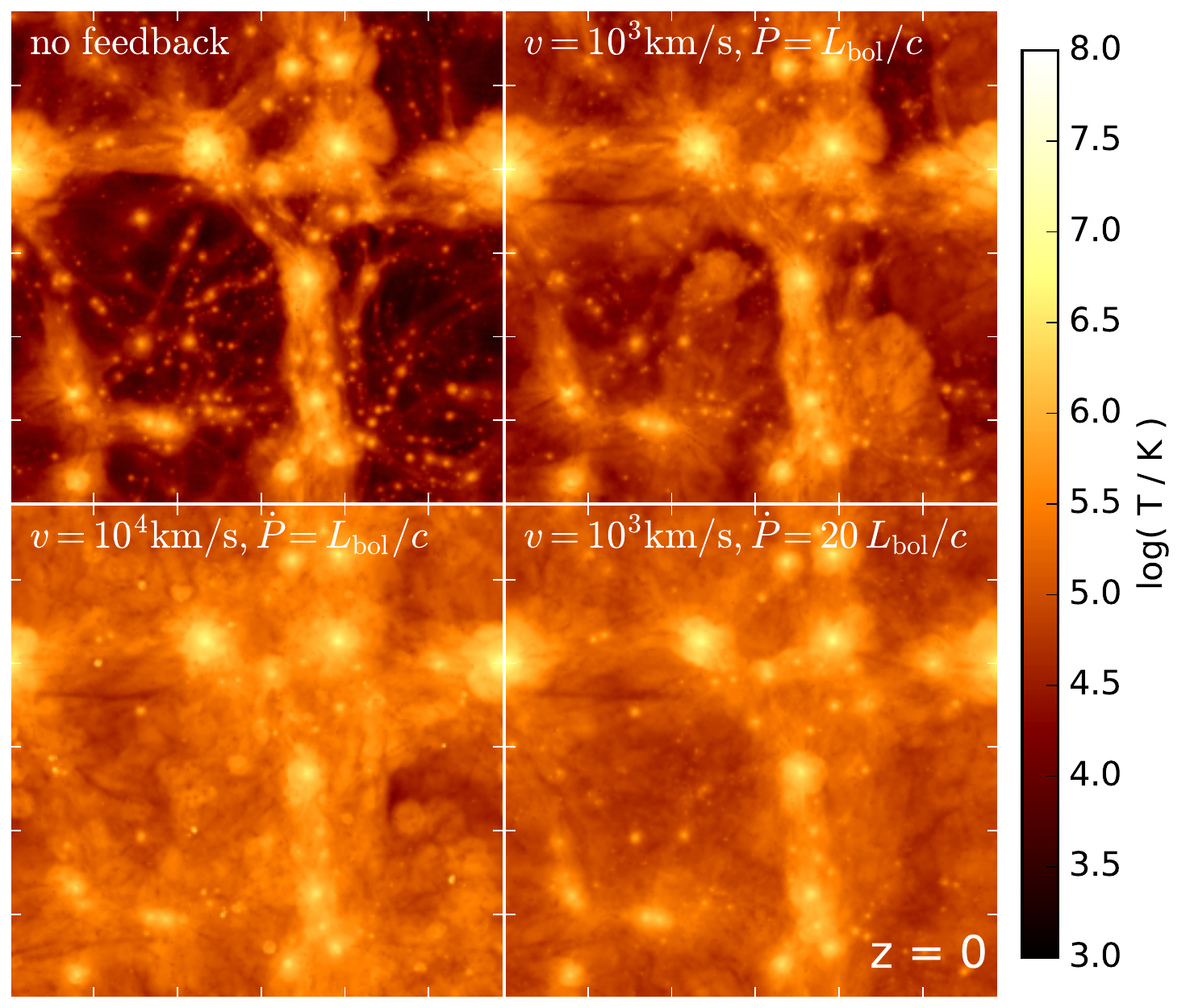}
\end{center}
\caption{Mass-weighted projected temperature distributions at $z=2$ (left) and $z=0$ (right) for simulations with different black hole feedback strengths. 
For each redshift, we compare simulations including (i) no black hole feedback (top left), (ii) $v_{\rm out} = 10^3$\,km\,s$^{-1}$ and $\dot{P}_{\rm out} = L_{\rm bol}/c$ (top right; fiducial simulation), (iii) $v_{\rm out} = 10^4$\,km\,s$^{-1}$ and $\dot{P}_{\rm out} = L_{\rm bol}/c$ (bottom left), and (iv)  $v_{\rm out} = 10^3$\,km\,s$^{-1}$ and $\dot{P}_{\rm out} = 20\,L_{\rm bol}/c$ (bottom right).
Each panel represents the full simulated volume, i.e. $20\,\hmpc$ comoving on a side.
Black hole feedback strongly affects the thermal properties of the intergalactic medium for the range of feedback parameters considered, including kinetic efficiencies $\epsilon_{\rm k}$ as low as $10^{-3}$ and momentum-loading $p_{\rm b}$ as low as 1.}
\label{fig:fig5}
\end{figure*}

\subsection{Effects of the accretion rate normalization}\label{sec:eps}

Figure~\ref{fig:fig3} shows the evolutionary tracks of black holes and galaxies in the \MM~plane for the 100 most massive systems in the simulated volume, where we compare simulations using different accretion rate normalization, $\epsilon_{\rm T} = 5$ (orange), $\epsilon_{\rm T} = 0.5$ (blue), and $\epsilon_{\rm T} = 0.05$ (purple), but otherwise identical parameters. 
Once $M_{\rm BH} >> M_{\rm seed}$, we see clearly in each case that $M_{\rm BH} \propto \epsilon_{\rm T}$ for a given $M_*$.  In other words, black holes converge onto the \MM~relation corresponding to a given $\epsilon_{\rm T}$.  After the initial transitory growth phase, $\epsilon_{\rm T}$ controls the normalization, but not the shape, of our predicted scaling relation \citep{Angles-Alcazar2013}.
This is expected, given $\dot{M}_{\rm BH} \propto \epsilon_{\rm T}$, if the linear effect of changing the normalization of $\dot{M}_{\rm Torque}$ dominates over non-linear effects of black hole feedback on $\sim$kpc scales.

\subsection{Effects of black hole feedback on the \MM~relation}\label{sec:feed}

Figure~\ref{fig:fig4} shows the \MM~relation obtained at $z=0$ for simulations adopting different black hole feedback parameters.  Although different feedback choices do change both $M_*$ and $M_{\rm BH}$ significantly (discussed further below), black holes and galaxies appear to move along the same \MM~relation.  The best power law fit to our fiducial \MM~relation (top right) is reproduced in all panels for comparison.  Going from the no-feedback simulation (top left) to our strongest feedback case ($v_{\rm out} = 10^3$\,km\,s$^{-1}$, $\dot{P}_{\rm out} = 20\,L_{\rm bol}/c$; bottom right), $M_{\rm BH}(M_*)$ decreases by only a factor $\sim 2$ for black holes in $M_{\star} \sim 10^{10}$\,\Msun~galaxies.  Overall, all simulations are in good agreement with the observed relation, suggesting that galaxy-scale black hole feedback does not play a primary role in establishing the scaling relations.

\section{Effects of black hole feedback}\label{sec:glob}

We now examine the impact of black hole feedback on properties other than the \MM~relation.

\subsection{Thermal properties of the intergalactic medium}

Figure~\ref{fig:fig5} illustrates the impact of black hole feedback on cosmological scales by showing the projected (mass-weighted) gas temperature distribution at $z=2$ and $z=0$.  As in Figure~\ref{fig:fig4}, we compare simulations with different feedback parameters.  In the absence of black hole feedback, the main heating sources of the intergalactic medium in our simulations are the photoionizing background and virial shocks that develop as gas accretes onto dark matter halos.

The impact of black hole feedback on the thermal properties of the intergalactic medium is evident already at $z=2$ in our fiducial simulation.  Black hole driven outflows create bubbles of hot gas expanding over scales significantly larger than the host dark matter halos.
The effects of black hole feedback on large scales become even more dramatic when we increase the outflow velocity to $v_{\rm out} = 10^4$\,km\,s$^{-1}$: a significant portion of the IGM is heated to temperatures $\gtrsim 10^6$\,K.  Similar large scale effects are seen for $v_{\rm out} = 10^3$\,km\,s$^{-1}$ and $\dot{P}_{\rm out} = 20\,L_{\rm bol}/c$.

\begin{figure}
\begin{center}
\includegraphics[scale=0.55]{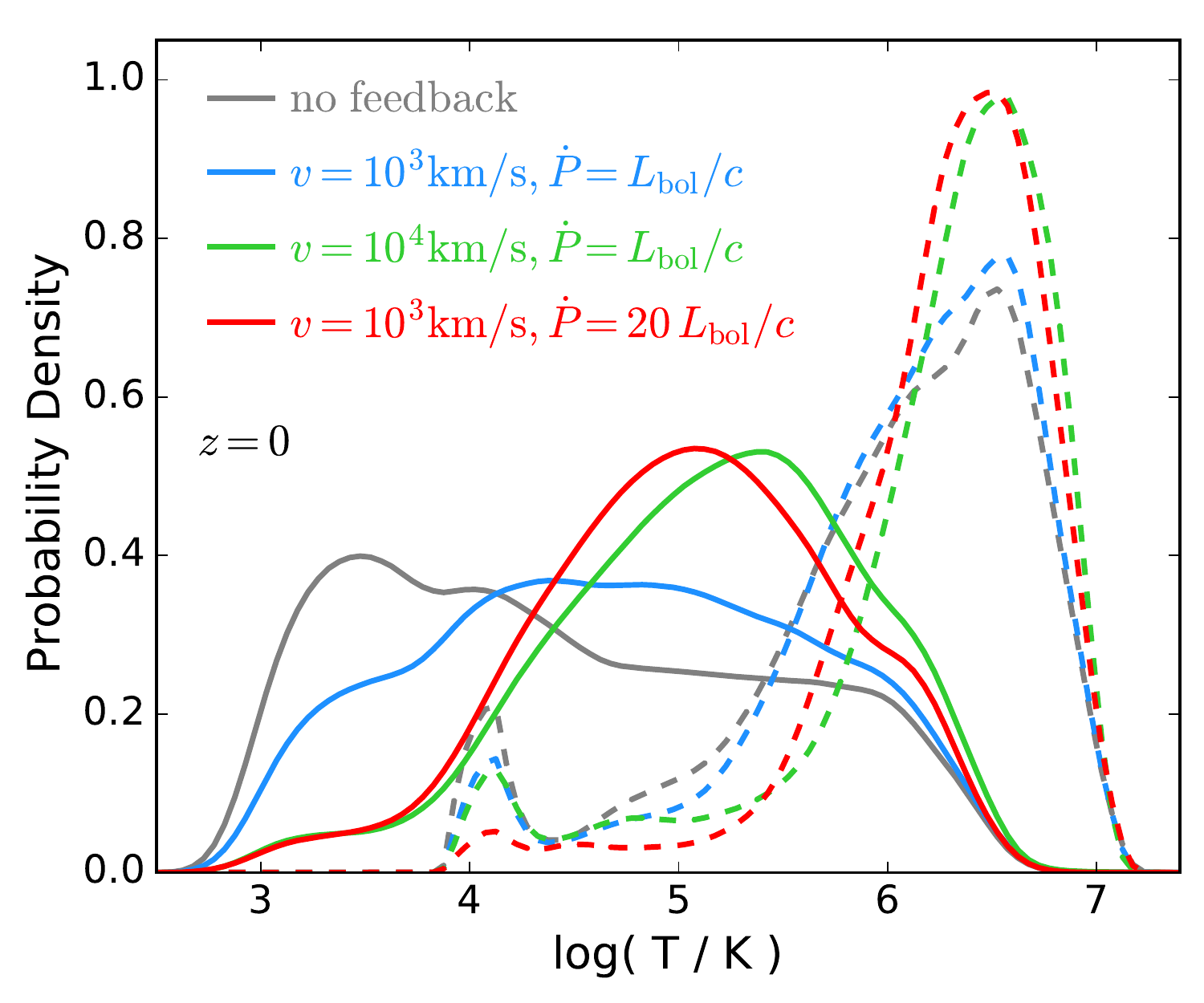}
\end{center}
\caption{Mass-weighted probability distribution for the temperature of the gas within halos (ignoring star-forming gas; dashed lines) and outside of halos (solid lines) at $z=0$ for simulations with different feedback parameters, including (i) no black hole feedback (gray), (ii) $v_{\rm out} = 10^3$\,km\,s$^{-1}$ and $\dot{P}_{\rm out} = L_{\rm bol}/c$ (blue; fiducial simulation), (iii) $v_{\rm out} = 10^4$\,km\,s$^{-1}$ and $\dot{P}_{\rm out} = L_{\rm bol}/c$ (green), and (iv)  $v_{\rm out} = 10^3$\,km\,s$^{-1}$ and $\dot{P}_{\rm out} = 20\,L_{\rm bol}/c$ (red).
Inside massive halos, virial shocks dominate heating and black hole feedback is a secondary effect.  Outside halos, fast black hole driven winds can increase the IGM temperature by factors $\sim$3--10.}
\label{fig:fig6}
\end{figure}

We quantify in Figure~\ref{fig:fig6} the efficiency of black hole feedback heating by showing mass-weighted temperature distributions for the gas within the virial radius of dark matter halos (dashed lines) and outside of halos (solid lines) at $z=0$.  Only gas particles with density below the threshold for star formation are included ($n_{\rm H} \lesssim 0.13$\,cm$^{-3}$).
The temperature distribution of gas within halos is shifted towards higher temperatures in simulations with black hole feedback, where the median temperature increases by $\sim 15$\,\%, 70\,\%, and 60\,\% relative to the no-feedback simulation for our fiducial, high velocity ($v_{\rm out} = 10^4$\,km\,s$^{-1}$), and large momentum boost ($\dot{P}_{\rm out} = 20\,L_{\rm bol}/c$) simulations, respectively.
As expected from Figure~\ref{fig:fig5}, the relative impact of black hole feedback on gas at larger scales is more prominent.  The median temperature of gas outside of halos increases by roughly one order of magnitude from no-feedback to our high velocity and large momentum boost simulations.

\begin{figure}
\begin{center}
\includegraphics[scale=0.55]{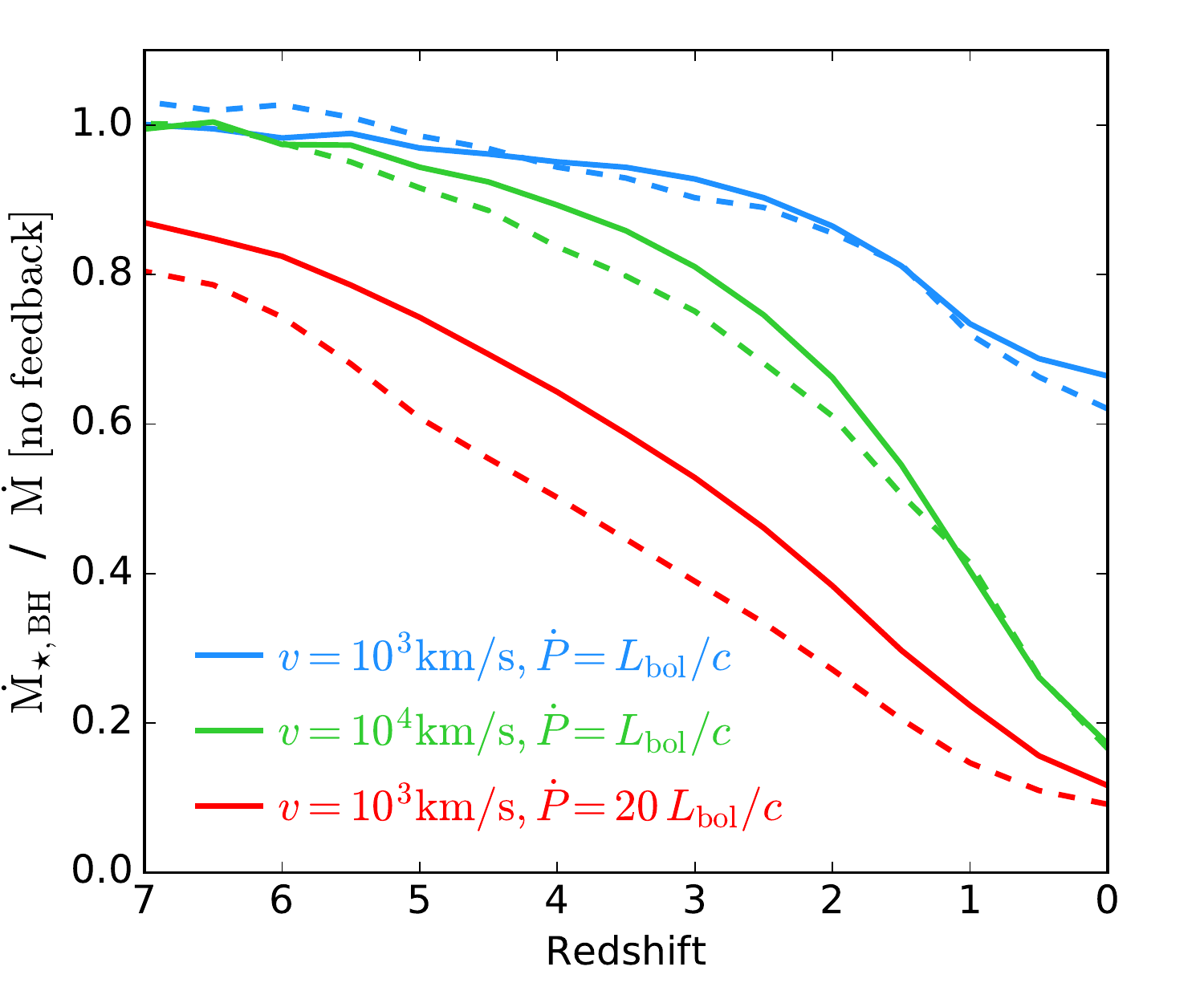}
\end{center}
\caption{Impact of galaxy-scale black hole feedback on the global growth of black holes and galaxies.  
We show the volume-integrated star formation rate (solid lines) and black hole growth rate (dashed lines), normalized to the corresponding rates in our no-feedback simulation.  We compare simulations with outflow velocity and total momentum flux (i) $v_{\rm out} = 10^3$\,km\,s$^{-1}$ and $\dot{P}_{\rm out} = L_{\rm bol}/c$ (blue; fiducial simulation), (ii) $v_{\rm out} = 10^4$\,km\,s$^{-1}$ and $\dot{P}_{\rm out} = L_{\rm bol}/c$ (green), and (iii)  $v_{\rm out} = 10^3$\,km\,s$^{-1}$ and $\dot{P}_{\rm out} = 20\,L_{\rm bol}/c$ (red).  Black hole driven outflows reduce the SFR and black hole accretion rate by a similar amount, increasingly so for simulations with higher feedback efficiency and lower redshift.}
\label{fig:fig7}
\end{figure}

\subsection{Stellar and black hole growth as a function of redshift}

Figure~\ref{fig:fig7} illustrates the effect of black hole feedback on the volume-integrated mass growth in stars and black holes at different redshifts.  Our fiducial feedback parameters yield a modest but systematic decrease in the total star formation and black hole growth rate relative to the no-feedback simulation, most noticeable at $z \lesssim 2$.  
The effects are weaker at high redshift because (1) the black holes are just seeded, and not yet massive, and (2) the galaxies are mostly low-mass, where black hole feedback is weak.
As expected, increasing the energy or momentum-loading of outflows increases these effects. For our high velocity ($v_{\rm out} = 10^4$\,km\,s$^{-1}$) and large momentum boost ($\dot{P}_{\rm out} = 20\,L_{\rm bol}/c$) simulations, the volume-integrated SFR and black hole accretion rate are reduced by a factor $\sim 5$--10 at $z=0$. 
Galaxy-scale AGN feedback suppresses global stellar and black hole growth by similar amounts, maintaining the slope of the \MM~relation but leading to the reduction in the number of systems at the high mass end in Figure~\ref{fig:fig4}.

\begin{figure*}
\begin{center}
\includegraphics[scale=0.55]{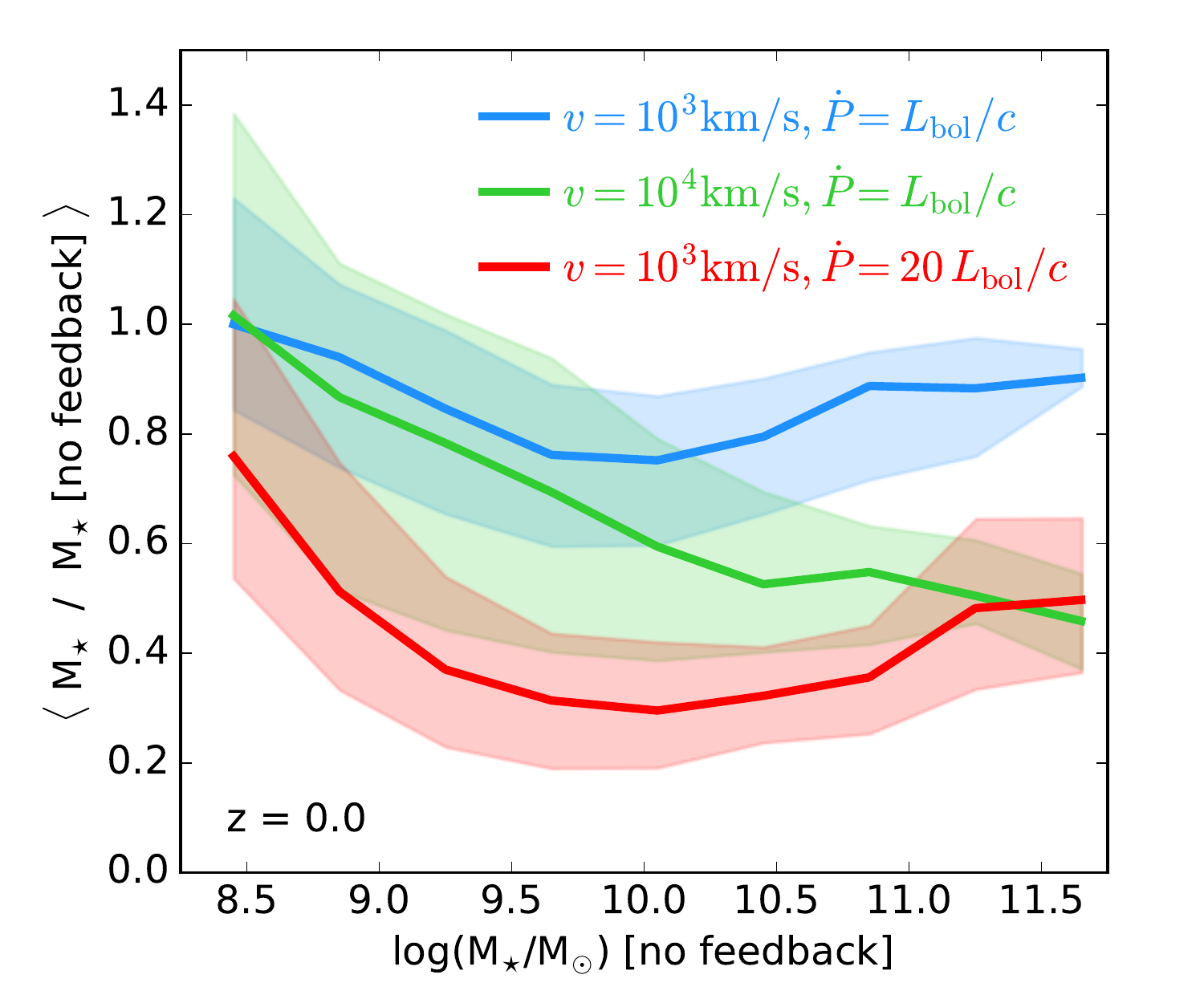}
\includegraphics[scale=0.55]{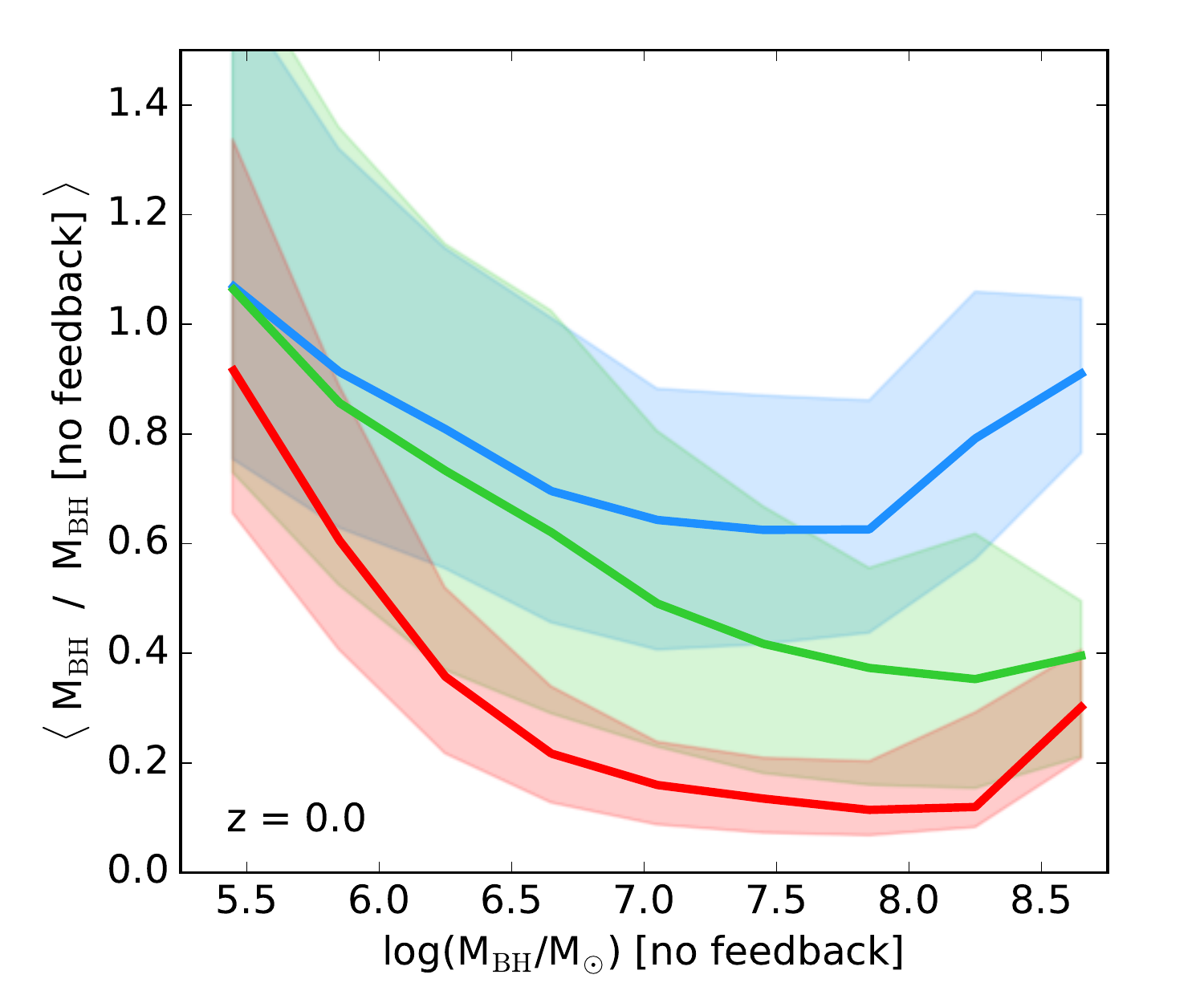}
\end{center}
\caption{Mass dependent effects of galaxy-scale black hole feedback. 
Left: ratio of the stellar mass of galaxies at $z=0$ in simulations including black hole feedback to the stellar mass of the corresponding galaxies in the no-feedback simulation as a function of stellar mass.  Different colors indicate simulations with different outflow velocity and momentum flux.  Solid lines and shaded regions indicate median and 10\,\%--90\,\% percentile ranges.
Right: same as left for black hole masses.
Black hole feedback appears to be more effective in galaxies with mass M$_{\star} \sim10^{10}$\,\Msun~for outflows with velocity $v_{\rm out} = 10^3$\,km\,s$^{-1}$, while the mass suppression efficiency increases to higher mass galaxies for simulations with $v_{\rm out} = 10^4$\,km\,s$^{-1}$.  The overall effect of feedback on the growth of black holes and galaxies is qualitatively similar.}
\label{fig:fig8}
\end{figure*}

\subsection{Galaxy mass dependence of black hole feedback effects}

Figure~\ref{fig:fig8} illustrates how the effects of black hole feedback depend on galaxy stellar mass.
We cross-match galaxies between simulations based on the unique ID of the particles they contain at $z=0$.  Each galaxy is matched to the galaxy that contains the largest number of common star particles, which we use to compute the stellar mass of galaxies in simulations with black hole feedback relative to the no-feedback simulation.

Black hole feedback in our simulations with $v_{\rm out} = 10^3$\,km\,s$^{-1}$ is more effective at suppressing star formation and black hole growth in galaxies in the mass range $M_{\star} \approx 10^{9}$--$10^{11}$\,\Msun~at $z=0$.
This trend with $M_{\star}$, already present at $z=2$, suggests that black holes in lower mass galaxies have not coupled enough energy to affect their host galaxies from the time of seeding down to $z=0$ (which may be sensitive to our seed model), while outflows with $v_{\rm out} = 10^3$\,km\,s$^{-1}$ may not be sufficient to unbind gas in galaxies at the highest masses.
With the same momentum flux and $v_{\rm out} = 10^4$\,km\,s$^{-1}$, the results are similar at low $M_*$ but the mass suppression efficiency keeps increasing toward high masses, giving a $\sim 60$\,\% reduction of the stellar mass of galaxies with $M_{\star} \gtrsim 10^{11.5}$\,\Msun~by $z=0$.  As expected, our simulation with $v_{\rm out} = 10^3$\,km\,s$^{-1}$ and  $\dot{P}_{\rm out} = 20\,L_{\rm bol}/c$ yields the strongest suppression of stellar growth at all masses, but the efficiency follows a trend with $M_{\star}$ similar to that of the fiducial simulation.  This suggests that the outflow velocity may have interesting implications for the suppression of galaxy growth at the highest masses.
Black holes suppress their own growth with increasing efficiency toward higher masses roughly in a similar way as they suppress the growth of their host galaxies.  Thus, the \MM~relation is roughly preserved.

\section{Discussion}\label{sec:dis}

We have implemented a self-consistent black hole growth model into the GIZMO code based on the analytic gravitational torque mass inflow rate of \citet{Hopkins2011_Analytic}.  This model captures the key scalings governing angular momentum transport from galactic scales down to parsec scales and reproduces the average gas inflow rates found in idealized nuclear scale simulations \citep{Hopkins2010_MultiScale,Hopkins2015_NuclearSims}.  
We have further implemented a kinetic black hole feedback model coupled to accretion.  This model does not attempt to explicitly capture wind driving mechanisms, but builds on previous kinetic feedback implementations in the literature applied to nuclear scale simulations \citep{Hopkins2015_NuclearSims}, galaxy merger simulations \citep{Choi2012_BHmodel,Debuhr2012}, and cosmological ``zoom-in" simulations \citep{Choi2015_CosmoSim}. 
In large volume cosmological simulations, we explore the effects of outflows with parameters similar to observed fast nuclear outflows \citep[$v_{\rm out} = 10^4$\,km\,s$^{-1}$, $\dot{P}_{\rm out} = L_{\rm bol}/c$; e.g.][]{Tombesi2013,Nardini2015} and galaxy-scale AGN-driven winds \citep[$v_{\rm out} = 10^3$\,km\,s$^{-1}$, $\dot{P}_{\rm out} = 20\,L_{\rm bol}/c$; e.g.][]{Faucher-Giguere2012_FeLoBAL,Faucher-Giguere2012_WindModel,Cicone2014,Harrison2014,Stern2015}.

Our simulations show that black hole feedback can have a large impact on the thermodynamic properties of the intergalactic medium as well as the overall growth of galaxies and massive black holes, in qualitative agreement with previous work \citep[e.g.][]{Sijacki2007,Vogelsberger2014,Schaye2015}.  Relative to our no-feedback simulation, black hole driven outflows yield a reduction in the total production of stars in the simulated volume by $z=0$ of $\sim 20$\,\%, 40\,\%, and 60\,\% for our fiducial, high velocity ($v_{\rm out} = 10^4$\,km\,s$^{-1}$), and large momentum boost ($\dot{P}_{\rm out} = 20\,L_{\rm bol}/c$) simulations, respectively.  
These correspond to momentum-loading factors $(1,1,20)\,L_{\rm bol}/c$ and energy-loading factors $(0.1,1,3)\%\,L_{\rm bol}$.  
Kinetic AGN outflows can thus have a large impact even for energetic efficiencies as low as $0.1\%\,L_{\rm bol}$ \citep{HopkinsElvis2010}.
Black holes suppress their own growth by similar, somewhat larger, factors, preserving the black hole--host scaling relations.

The gravitational torque accretion model has several important consequences. Because the inflow rate from this mechanism is approximately proportional to the nuclear gas supply, a linear black hole--host mass scaling emerges naturally (with slope and scatter in good agreement with that observed), {\em independent} of galaxy-scale black hole feedback (coupled on $\sim\,$kpc scales).  Although black hole feedback on these large scales does suppress both black hole growth and galaxy growth, removing gas from large-scale reservoirs suppresses both by a similar amount, moving systems along (not off) the scaling relations. In short, black hole and central galaxy mass are determined by a common gas supply modulated by gravitational torques, as increasingly suggested by observations of AGN in star forming galaxies \citep[e.g.][]{Rafferty2011,Mullaney2012_AGN_MainSequence, Chen2013,Rosario2013_AGNinSFGals,Heckman2014,Hickox2014,Vito2014_AGNgasrich,Dai2015,Delvecchio2015,Sabater2015,Sun2015,Trump2015}.  
Because the black hole fueling rate is determined by gravitational instabilities and resulting torques, it is nearly independent of black hole mass.  This in turn means that the black hole--host scaling relations are insensitive to the ``seed'' black hole mass.   
Under-massive and over-massive black holes grow proportionally faster and slower than their host galaxies, respectively, converging onto the scaling relations without the need for self-regulation by galaxy-scale feedback \citep{Angles-Alcazar2013}.
Similar convergence may be indicated by recent observations of accreting black holes in star forming galaxies at $z \lesssim 2$ (\citealp{Sun2015}; see also \citealp{Merloni2010_ScalingEvol}).  Merging of galaxies and central black holes may help reduce the scatter of the scaling relations \citep{Peng2007,Hirschmann2010,Jahnke2011} but we find it is not a significant contribution \citep{Angles-Alcazar2015}.

In our simulations, the normalization of the black hole--host relation is controlled by the normalization of the mean accretion rate from $\sim$\,kpc scales down to the black hole ($\epsilon_{\rm T}$). To match the $z=0$ scaling, we require a factor $\sim 10$ lower $\epsilon_{\rm T}$ than the inflow rate down to $\sim$\,pc scales estimated in \citet{Hopkins2011_Analytic}.  This discrepancy can be interpreted as mass loss in winds from the accretion disk, suppressing black hole accretion relative to the inflowing gas driven by gravitational torques \citep{Angles-Alcazar2013,Angles-Alcazar2015}.
Moreover, $\epsilon_{\rm T}$ can be modified by stellar and black hole feedback on small scales (below those resolved here), which were not modeled explicitly in \citet{Hopkins2011_Analytic}.   
More recent simulations in \citet{Hopkins2015_NuclearSims} showed that accretion-disk winds coupling to the gas on small scales can vent some hot, fast material to large scales (the galaxy-scale black hole feedback modeled here), while driving slower outflows in the cold, dense gas forming the high-column density (e.g. torus) regions at $\lesssim 100\,$pc. The latter do not escape the galaxy center, but can suppress accretion by a factor $\sim10$ by evacuating gas from the vicinity of the black hole. In this sense, black hole feedback on small scales may still play a significant role determining the normalization of the scaling relations. 
In our models, $\epsilon_{\rm T}$ is independent of redshift, which yields redshift-independent black hole--host correlations.  However, given systematic redshift evolution in average Eddington ratios \citep{Angles-Alcazar2015} and the typical densities, metallicities, and star formation rate properties in galactic nuclei, it is plausible that $\epsilon_{\rm T}$ (hence the black hole--host scalings) could evolve \citep[see, e.g.][]{Hopkins2007_BHplane,DiMatteo2008,DeGraf2015,Sijacki2015}. Observations remain inconclusive regarding such evolution \citep[e.g.][]{Bongiorno2014,Schulze2014,Shen2015,Sun2015,Willott2015}.

In this study, we have deliberately simplified the complexity of galaxy formation physics in our simulations to isolate the effects of black holes and make large volume simulations more feasible.
Our simulations utilize the sub-grid model of \citet{Springel2003_Multiphase} instead of resolving a multi-phase ISM and do not include stellar-feedback driven winds \citep[e.g.][]{Dave2011_GasMet,Dave2011_MstarSFR,Agertz2013,Angles-Alcazar2014,Hopkins2014_FIRE}.  
The impact of black hole feedback at galactic scales may depend on these properties \citep[e.g.][]{Gabor2014,Hopkins2015_NuclearSims,Roos2015}.
Future work will need to consider a more detailed galaxy formation model. 
Finally, while the physics included in the gravitational torque model is well motivated and predicts the inflow rates measured in nuclear scale simulations significantly better than other models \citep{Hopkins2011_Analytic,Hopkins2015_NuclearSims}, additional mechanisms for angular momentum transport (e.g. scattering of dense gas clumps and gravitational instability-driven turbulence) should be considered in regimes where it may not be appropriate.

\section{Conclusions}\label{sec:conc}

Modeling black hole growth and feedback in a cosmological context continues to be a significant challenge even in the latest cosmological hydrodynamical simulations.  The models presented here emphasize (1) the importance of gravitational torques regulating a common gas supply for star formation and black hole growth and (2) the potential impact of AGN-driven outflows on galaxy evolution.  
Our results suggest that the efficiency with which gravitational torques feed the central black hole relative to the host galaxy star formation rate play a primary role on the observed connection between massive black holes and galaxies, while the scaling relations are relatively insensitive to the amount of black hole feedback injected at galactic scales.  
This highlights the importance of using observations other than the scaling relations to constrain black hole feedback models, including direct measurements of outflow properties and the thermodynamic state of gas in the intergalactic medium.  
In future work, we will extend this study to higher resolution simulations with more realistic ISM physics to investigate whether our main conclusions continue to hold as physical processes operating below the resolution of our present cosmological simulations are explicitly resolved.

\acknowledgments

We thank M. Elitzur, E. Quataert, and P. Torrey for useful discussions and M. van Daalen for providing the initial conditions.
DAA acknowledges support by a CIERA Postdoctoral Fellowship.
RD acknowledges support from the South African Research Chairs Initiative and the South African National Research Foundation, and funding from NASA ATP grant NNX12AH86G to the University of Arizona.
CAFG was supported by NSF through grants AST-1412836 and AST-1517491, by NASA through grant NNX15AB22G, and by STScI through grants HST-AR-14293.001-A and HST-GO-14268.022-A.
F{\"O} acknowledges support from NSF grant AST-1108753 and NASA TCAN award NNX14AB48G.
Support for PFH was provided by an Alfred P. Sloan Research Fellowship, NASA ATP Grant NNX14AH35G, and NSF Collaborative Research Grant \#1411920 and CAREER grant \#1455342. 
Numerical calculations were run using Northwestern University's compute cluster ``Quest" and the Extreme Science and Engineering Discovery Environment (XSEDE), which is supported by NSF grant ACI-1053575.
This work benefited from the hospitality of the Aspen Center for Physics, supported by NSF grant PHY-1066293.

\vspace{0.4cm}

\bibliography{bhs_mnras}

\begin{appendix}

\section{Numerical Robustness}\label{sec:appendix:rob}

Different aspects of the numerical robustness of our gravitational torque accretion model have been discussed in \citet{Angles-Alcazar2013,Angles-Alcazar2015}.  These include numerical convergence properties as well as uncertainties associated with the bulge-disk decomposition method and the size of the radial aperture used in the evaluation of equation~(\ref{eq:torque}). 
Here, we present a resolution convergence test of our coupled accretion and feedback model, evaluate the dependence of our results on the hydrodynamics solver employed, and explore the implications of different definitions of host galaxy bulge mass on the simulated \MM~relation.

\subsection{Resolution convergence}\label{sec:appendix:res}

In order to test our results for numerical convergence, we replicate our fiducial simulation with $8\times$ higher mass resolution and $2\times$ higher force resolution by evolving $2 \times 512^3$ particles down to $z = 2$ in a $[20\,\hmpc]^3$ comoving volume.  With the exception of the force softening length, all model parameters are identical to that of our fiducial simulation with $2 \times 256^3$ particles, including the number of neighbors used in the black hole accretion and feedback parameterization.   

\begin{figure}
\begin{center}
\includegraphics[scale=0.5]{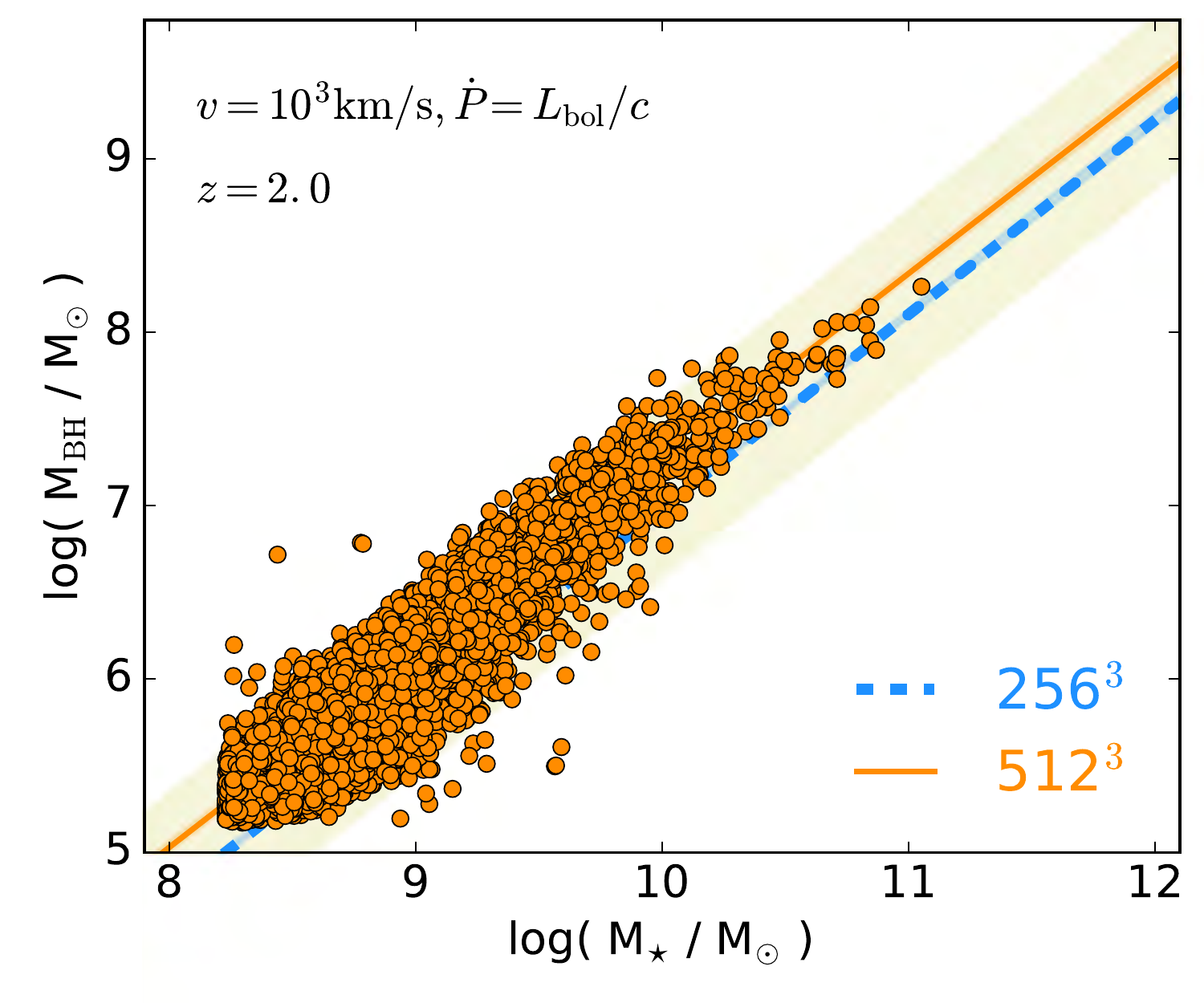}
\end{center}
\caption{\MM~relation at $z=2$ obtained for a simulation with $8\times$ higher mass resolution relative to our fiducial simulation, i.e. using $512^3$ gas and dark matter particles, but otherwise identical black hole accretion and feedback parameters. 
The orange solid line indicates the best power-law fit to the \MM~relation for the high resolution simulation, while the blue dashed line shows to the best fit relation for our fiducial simulation.  
The beige shaded area corresponds to 0.5 dex scatter in $M_{\rm BH}$ relative to \citet{Haring2004}. 
The increased resolution yields a very similar \MM~relation.}
\label{fig:figA1}
\end{figure}

Figure~\ref{fig:figA1} shows the \MM~relation at $z=2$ corresponding to our high resolution simulation.  The best power-law fit for galaxies with $M_{\star} > 10^{9.5}$\,\Msun, indicated by the orange solid line, is in good agreement with the local observed \Mbulge~relation of \citet{Haring2004}, though with slightly higher normalization.  Indeed, compared to the best fit relation for our fiducial simulation at $z=2$ (blue dashed line), the increased resolution yields a very similar slope but $\sim 0.25$ dex higher normalization.  
Lower gravitational resolution may result in the underestimation of disk fractions 
with the consequent reduction in black hole accretion rates relative to higher resolution simulations \citep{Angles-Alcazar2015}.
The numerical resolution tests presented in \citet{Angles-Alcazar2013} showed very good convergence of the \MM~relation between simulations $5\times$ and $40\times$ higher mass resolution relative to our fiducial simulation.  This suggests that the simulations presented here may have not yet reached full numerical convergence relative to the black hole accretion model.
Alternatively, a difference in the normalization of the \MM~relation for simulations with different resolution could partially arise owing to higher efficiency of feedback in simulations with lower resolution \citep{Bourne2015}.  
Nonetheless, while the normalization of the \MM~relation is mildly sensitive to resolution, the overall trends described in this work are unchanged.  
A modest re-normalization of $\epsilon_{\rm T}$ could compensate for the differences in resolution and our general conclusions are unaffected by this choice.

\subsection{Hydrodynamics solver}\label{sec:appendix:mfm}

\begin{figure}
\begin{center}
\includegraphics[scale=0.5]{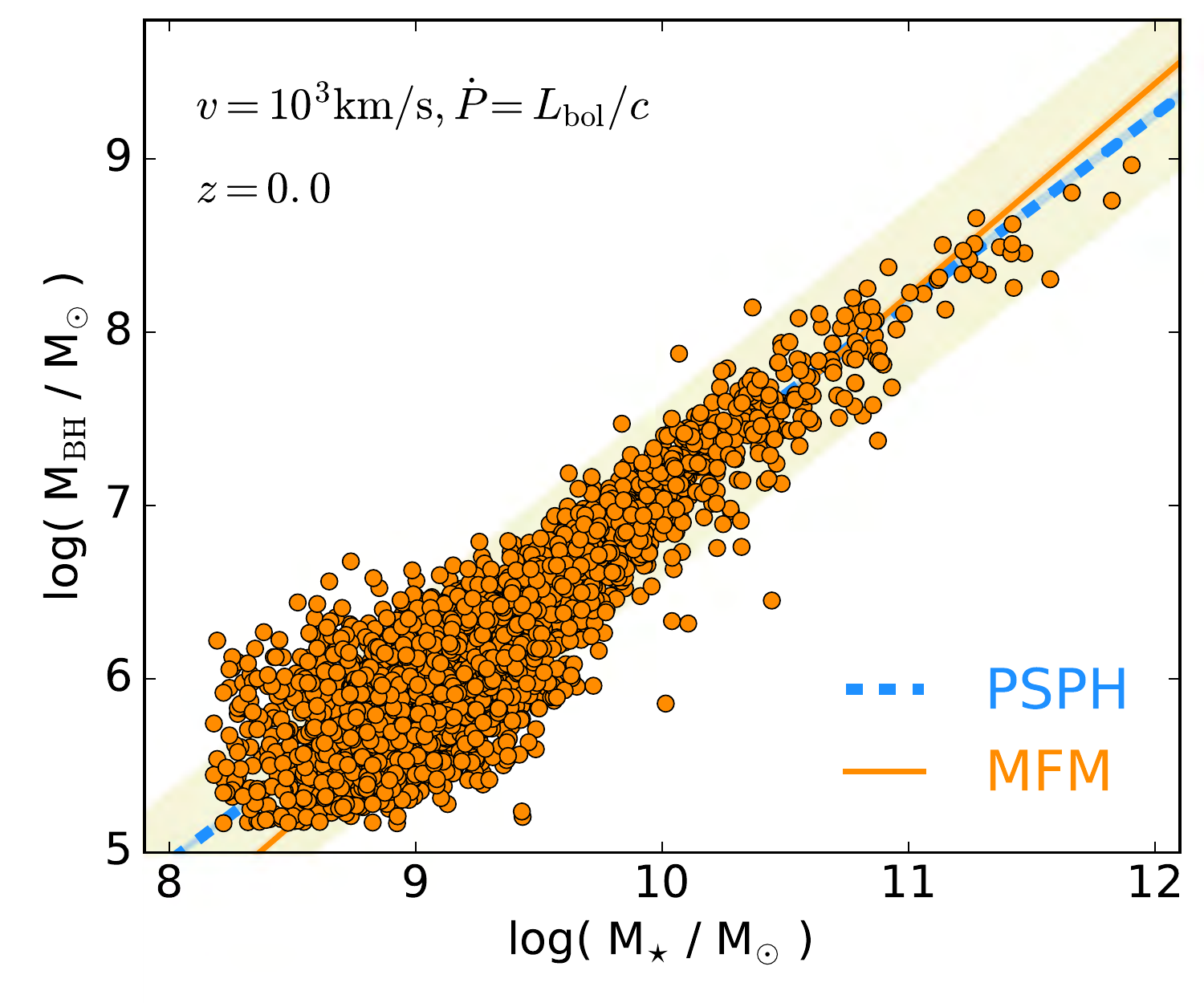}
\end{center}
\caption{\MM~relation at $z=0$ obtained for a simulation using the Godunov-type ``meshless finite mass" (MFM) hydrodynamics solver in GIZMO but otherwise identical black hole accretion and feedback parameters as our fiducial simulation (which used PSPH, the pressure formulation of smooth particle hydrodynamics). 
The orange solid line indicates the best power-law fit to the \MM~relation for the MFM simulation, while the blue dashed line shows to the best fit relation for our fiducial PSPH simulation.  
The beige shaded area corresponds to 0.5 dex scatter in $M_{\rm BH}$ relative to \citet{Haring2004}.
MFM and PSPH agree well: our uncertainties are not driven by the hydrodynamic method.}
\label{fig:figA2}
\end{figure}

We take advantage of the multi-method nature of the GIZMO code to evaluate the robustness of our results with respect to the hydrodynamics solver.  In particular, we compare our results using a pressure-entropy formulation of smooth particle hydrodynamics with the Lagrangian Godunov-type ``meshless finite mass" (MFM) method implemented in GIZMO \citep{Hopkins2015_Gizmo}.
Figure~\ref{fig:figA2} shows the \MM~relation at $z=0$ resulting from a simulation using the MFM hydrodynamics solver to evolve $256^3$ gas resolution elements in a $[20\,\hmpc]^3$ comoving volume.  The initial conditions and model parameters are identical to that of our fiducial simulation.  The only exception is the use of a cubic spline kernel with 32 neighbors instead of the quintic spline kernel with 64 neighbors used in our SPH simulations (since MFM converges at lower neighbor number).  Nonetheless, 
to preserve the physical scale at which black hole accretion and feedback are evaluated, the number of neighbors used for the black hole accretion and feedback prescriptions in the MFM simulation is the same as in the SPH simulations ($\sim256$ particles).

The $z=0$ \MM~relation obtained with MFM is in very good agreement with the observed \Mbulge~relation, with most black hole--galaxy pairs located within 0.5 dex of the \citet{Haring2004} relation.  Compared to our fiducial SPH simulation (Figure~\ref{fig:fig1}), MFM produces slightly larger scatter in the low mass regime. Considering the best power-law fit to the \MM~relation for galaxies with $M_{\star} > 10^{9.5}$\,\Msun~(indicated by the orange solid line), MFM yields slightly steeper slope and lower normalization relative to our fiducial simulation (blue dashed line).  
Overall, the good agreement between the two hydrodynamic methods, with no additional calibration of model parameters, confirms that our conclusions are not sensitive to the choice of hydrodynamics solver.

\begin{figure*}
\begin{center}
\includegraphics[scale=0.5]{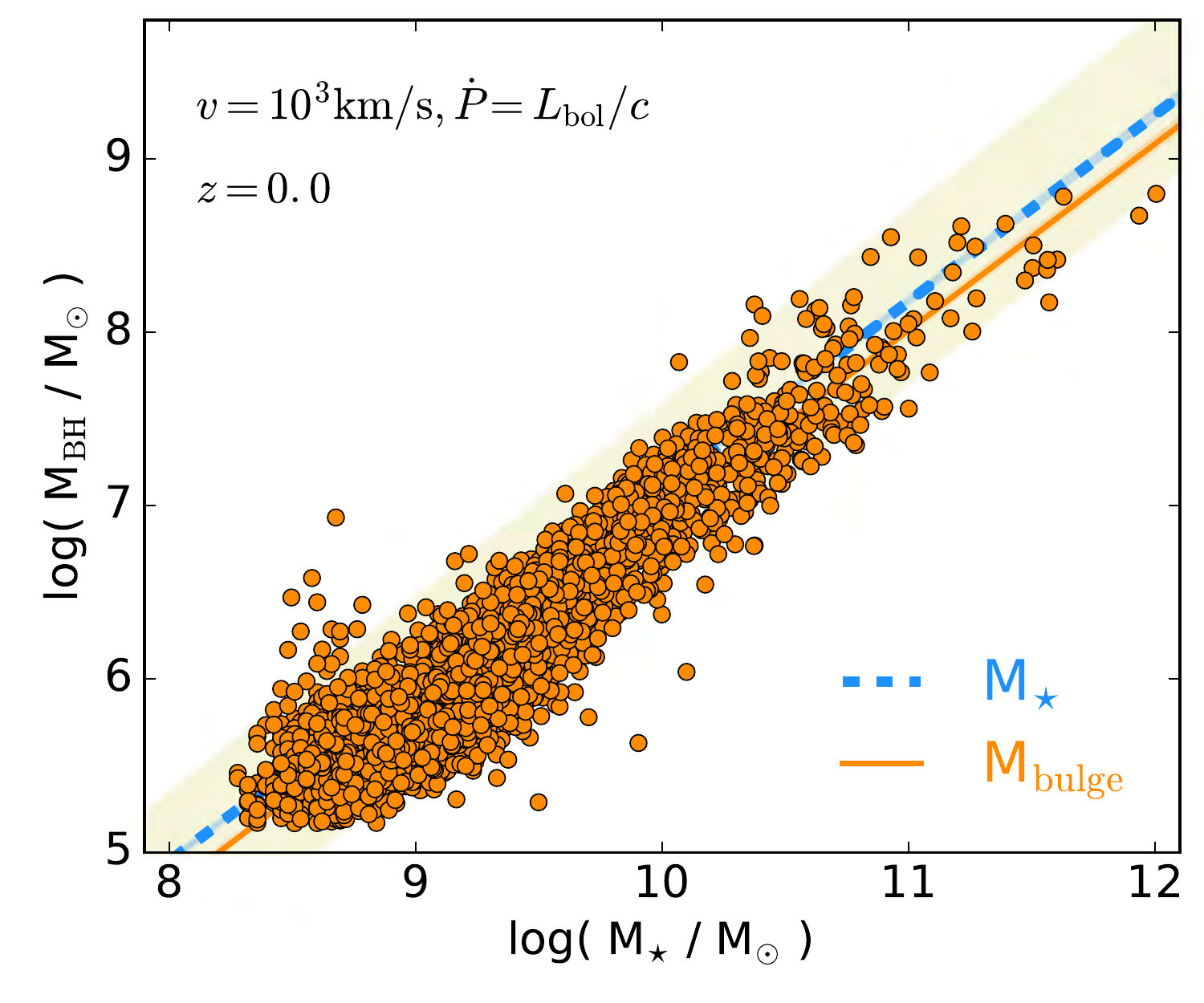}
\includegraphics[scale=0.5]{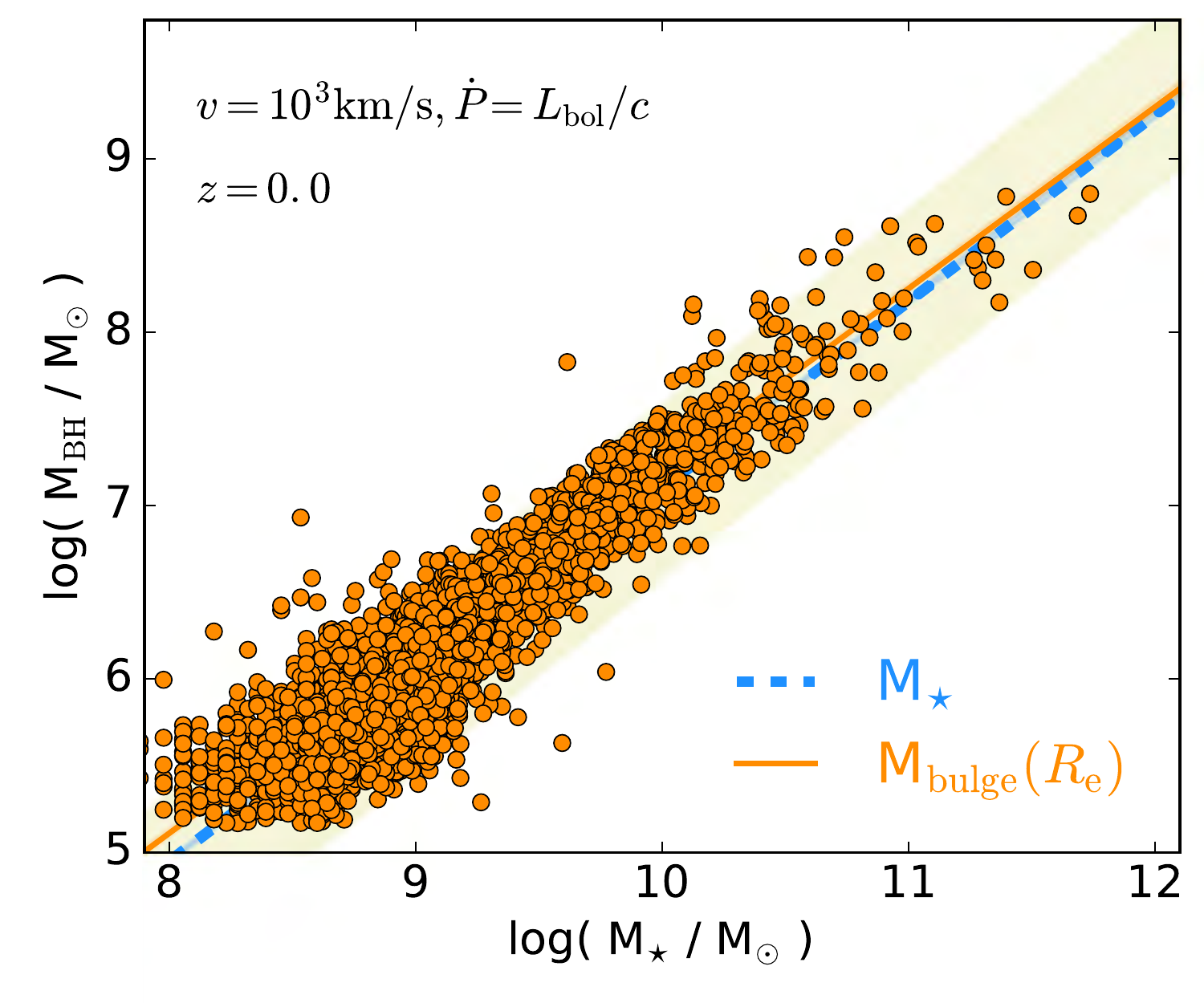}
\end{center}
\caption{\MM~relation at $z=0$ for our fiducial simulation using different definitions of host galaxy bulge mass.  Left: $M_{\rm BH}$ is plotted against the total mass of the spheroidal component computed using a full three-dimensional kinematic decomposition.  Right: same as the left panel for the bulge mass computed within the effective radius of the galaxy.
The orange solid lines indicate the best power-law fit to the \MM~relation for the new $M_{\star} \equiv M_{\rm bulge}$ definitions.  The blue dashed lines show to the best fit relation for our standard definition, where $M_{\star}$ is computed as the total stellar mass within the effective radius of the galaxy.  
The beige shaded area corresponds to 0.5 dex scatter in $M_{\rm BH}$ relative to \citet{Haring2004}.
Similar scaling relations are obtained regardless of the exact definition of $M_{\rm bulge}$ employed in our large-volume cosmological simulations with relatively limited resolution.}
\label{fig:figA3}
\end{figure*}

\subsection{Bulge-disk decomposition}\label{sec:appendix:mbulge}

Throughout this paper, the stellar mass within the effective radius of the galaxy ($M_{\star}$) has been used as proxy for bulge mass when comparing simulation results with the observed \Mbulge~relation.  This type of simplification is commonly used in cosmological simulations \citep[e.g.][]{DeGraf2015,Sijacki2015} and 
is justified by the fact that bulge-disk decompositions are very uncertain at the resolutions achieved in typical cosmological simulations.  
Indeed, producing galaxies with realistic bulges continues to be a challenge even for high resolution cosmological ``zoom-in" simulations \citep{Brooks2015_BulgeRev}.  
In contrast, numerical convergence for the stellar mass of galaxies in cosmological simulations is significantly better than any estimate of the bulge component, allowing for a simple but robust quantification of the relative growth of black holes and galaxies.  In addition, using $M_{\star}$ facilitates comparisons with observational studies at higher redshifts, where bulge masses are difficult to estimate \citep[e.g.][]{Jahnke2009,Sun2015}.
Nonetheless, it is important to address the implications of different definitions of host galaxy bulge mass on the simulated scaling relation within the limitations of the numerical resolution.

Figure~\ref{fig:figA3} shows the \MM~relation at $z=0$ for our fiducial simulation using two different definitions of host galaxy bulge mass.  We perform a simple bulge-disk kinematic decomposition using the full three-dimensional information available in the simulation.  For each galaxy, we compute the angular momentum vector of the stellar component, which is used as the reference axis to calculate the azimuthal velocity ($v_{\phi}$) of each star particle. The mass of the spheroidal component ($M_{\rm bulge}$) is estimated as double the mass of particles moving with $v_{\phi} < 0$ \citep{Abadi2003,Angles-Alcazar2014}.  The left panel of Figure~\ref{fig:figA3} shows the \MM~relation for the total spheroidal component.  On average, $M_{\rm bulge}$ is larger than the stellar mass within the effective radius, which yields $\sim 0.16$ dex lower normalization in the best fit relation relative to the best power-law fit using our standard definition of $M_{\star}$.  The right panel shows the \MM~relation for the spheroidal component computed only for star particles within the effective radius of the galaxy, $M_{\rm bulge}(R_{\rm e})$.  In this case, $M_{\rm bulge}(R_{\rm e}) \leq M_{\star}$ and the best power-law fit yields $\sim 0.10$ dex higher normalization and slightly lower slope compared to our standard definition of the \MM~relation.  In either case, our results depend only weakly on the exact definition of the bulge component and our main conclusions remain unchanged.  
Nonetheless, higher resolution zoom-in simulations \citep[e.g.][]{Hopkins2014_FIRE} will be necessary to properly address whether our black hole accretion and feedback model predicts different correlations between $M_{\rm BH}$ and various galaxy components.

\end{appendix}

\end{document}